# Active Learning for Generalizable Detonation Performance Prediction of Energetic Materials

R. Seaton Ullberg[1,*], Megan C. Davis[1], Jeremy N. Schroeder[1,2], Andrew H. Salij[1], M. J. Cawkwell[1], Christopher J. Snyder[3], Wilton J. M. Kort-Kamp[1], Ivana Matanovic[1,*]

[1]Theoretical Division, Los Alamos National Laboratory, Los Alamos, NM 87545, USA

[2]Department of Mechanical Engineering, Texas Tech University, Lubbock, TX 79409, USA

[3]Weapon Stockpile Modernization Division, Los Alamos National Laboratory, Los Alamos, NM 87545, USA

**Abstract**

The discovery of new energetic materials is critical for advancing technologies from defense to private industry. However, experimental approaches remain slow and expensive while computational alternatives require accurate material property inputs that are often costly to obtain, limiting their ability to efficiently predict detonation performance across a vast chemical space. We address this challenge through an active learning strategy that integrates density functional theory calculations, thermochemical modeling, message-passing neural networks, and Bayesian optimization. The resulting high-throughput workflow iteratively expands the training dataset by selecting new molecules in a targeted manner that balances the exploration of broad chemical space with the exploitation of promising high-performing candidates. This approach yields the largest publicly available database of potential CHNO explosives drawn from an initial pool of more than 70 billion candidates and a generalizable surrogate model capable of accurately predicting detonation performance ($R^2 > 0.98$). Feature importance analysis on this largest-to-date dataset reveals that oxygen balance is the dominant driver of detonation performance, complemented by contributions from local electronic structure, density, and the presence of specific functional groups. Cheminformatics analysis highlights how energetic materials with similar performance metrics tend to cluster in distinct chemical spaces offering a clearer direction for future synthesis studies. Together, the surrogate model, database, and resulting chemical insights provide a valuable foundation for high-throughput screening and targeted discovery of new energetic materials spanning diverse and previously unexplored regions of chemical space.

*Corresponding authors: sullberg@lanl.gov, ivana@lanl.gov

## 1. Introduction

Energetic materials (EMs) are essential to a wide range of civilian and military applications, from mining and construction to their use as critical components in modern weapons systems.[1] Their significant utility and numerous constraints regarding performance, safety, and cost have promoted research and investment into the development of new EMs. Despite the many molecular EMs that have been synthesized as a result of this effort[2], the most commonly used EMs today were developed prior to World War II and present health hazards that need to be improved upon.[3] For example, both 2,4,6-trinitrotoluene (TNT) and 1,3,5-trinitroperhydro-1,3,5-triazine (RDX) are carcinogens with a synthesis procedure that results in the release of $NO_x$ fumes and acidic waste.[4–6] While there has been some success in addressing these environmental concerns with the development of safer EMs such as 3,3'-diamino-4,4'-azoxyfurazan (DAAF)[7] and 4H,8H-difurazano[3,4-b:3',4'-*e*]pyrazine (DFP)[8], improvements in detonation performance have largely stalled since the discovery of hexanitrohexaazaisowurtzitane (CL-20) in the 1980s.[9] Furthermore, the timeline of CL-20's development spans decades - with 15 years between the initial synthesis and initial production preceding another 15 years before full commercialization was realized.[9,10] Such a slow pace of development underscores the limitations of traditional discovery workflows and highlights the value of new methods capable of rapidly screening large numbers of candidate EMs. By using an accurate and efficient surrogate to evaluate detonation performance at scale, high-performing compounds can be identified and prioritized for development, while also enabling the discovery of materials with beneficial secondary properties including improved thermal stability, reduced handling sensitivity, and enhanced environmental health and safety. Moreover, this capability provides a natural foundation for guiding targeted molecular generative models toward high-performing regions of chemical space, expanding the EM design space beyond a limited set of historically dominant materials without compromising the primary performance-driven objectives.[11]

Recently, there has been a surge in the popularity of applying machine learning (ML) and artificial intelligence (AI) methods to solve problems in fields such as chemistry[12], biology[13], and materials science.[14] Improved access to high-performance computing resources and large databases has proven to be invaluable in developing accelerated computational approaches to drug discovery[15], materials property prediction[16], and synthetic pathway planning.[17] Considering the vast chemical search space within which new EMs might exist, predictive ML models present an exciting opportunity to accelerate the pace of new molecule discovery relative to experimental trial-and-error driven by chemical intuition or even high-throughput quantum mechanical calculations.

Machine learning models to predict the properties of EMs have been developed previously.[18–28] However, each implementation differs significantly in architecture, selection of input features, and the size and diversity of its training dataset. Some authors[18,19,21] have found success in using simple regression models such as kernel ridge regression (KRR)[29] or support vector machines (SVM)[30] while others[22,27,28] have resorted to more advanced neural network models to expose highly non-linear relationships in the training data. The choice of descriptors – numerical features that embed the composition and structure of a molecule into a unified input parameter for the model – is dictated by tradeoffs between the time required to compute them and the richness of the information they contain.[31] Expensive descriptors might be derived from experimental measurements or quantum mechanical calculations as is the case in the work of Yang et al.[27] and Casey et al.[22] Topological descriptors, which can be calculated directly from the molecular graph often represented as a Simplified Molecular Input Line Entry System (SMILES)[32], have the benefit of rapid computation time making them an ideal choice for high-throughput screening methodologies. Models trained on these descriptors exhibit similar or better performance to models trained on electronic structure derived properties indicating that the additional information afforded by quantum mechanical calculations may not be necessary to predict certain properties of EMs.[18,23,28,33] In addition to providing convenient topological descriptors, the graph representation of a molecule is naturally compatible with modern neural network architectures, particularly message passing neural networks (MPNNs), which can learn high-dimensional molecular embeddings that serve as rich descriptors for property prediction tasks.[34] This strategy has proven highly effective for developing accurate EM surrogate models[28,35–37], though it introduces another tradeoff: the resulting embedded feature space sacrifices physical interpretability. While interpretability metrics can be applied to graph-based models, their usefulness is largely limited to identifying specific subgraphs with outsized impact rather than revealing important higher-level features with obvious physical analogs.[38]

Dataset size and composition can also be a differentiating factor with some works using training data that includes fewer than 250 molecules[18,19,23] while others use more than 20,000.[20,22]

Larger and more chemically diverse datasets tend to be more difficult for a model to learn, but they provide the benefit of generalizing more effectively to molecules which are outside of the training set.[39]

A detonation in organic explosives involves the propagation of a strong shock wave that compresses and heats the molecules such that they undergo rapid exothermic reactions that produce small molecule product species. High pressure and high temperature conditions generated by the reaction sustain the propagation of the shock front. Detonation performance, typically measured as detonation velocity (the speed at which the shock front travels) or detonation pressure (the pressure measured at the shock front), is fundamental in characterizing the usefulness of an EM. Therefore, these metrics have become high-value targets to predict in the simulation and modeling community.[40] Some of the most significant advancements in this area have come from the work of Casey et al.[22] and Davis et al.[20] Both authors apply state-of-the-art modeling techniques to large EM datasets (approximately 20,000 molecules) and achieve predictive accuracy within their respective test sets.

Casey et al. trained a convolutional neural network on a 4D tensor of charge density and electrostatic potential over the entire spatial extent of a molecule. This high-dimensional descriptor requires an expensive electronic structure calculation for each molecule, but the information it provides offers a robust and physically meaningful representation for the model to learn from. However, this descriptor has the disadvantage that it is not rotationally invariant and thus all unique rotations around the three spatial axes (24 in total) must be included during training. This type of data augmentation approximates rotational invariance, but it drastically increases the amount of data required to train an effective model.

Davis et al. take an alternative approach, combining easily computed topological descriptors with electronic structure information obtained via a computationally efficient density functional tight binding method. They trained multiple ML models with different architectures and found that incorporating both descriptor types consistently but modestly improved predictive performance relative to models trained solely on topological features.

While previous studies present useful frameworks for rapid EM screening, most rely on training and test data drawn from a single database, either GDB[41–44] or the Cambridge Structural Database (CSD).[45] While both datasets contain thousands of molecules with energetic characteristics, such as known explosophores, this reliance on a single data source limits model generalizability, particularly when extrapolating to diverse regions of chemical space that are of special interest in materials discovery campaigns.

In this work, we approach the challenge of predicting detonation performance with a novel active learning (AL) strategy designed to maximize the generalizability of the final model across chemical space while minimizing the amount of training data required to do so. Although AL has been applied previously to predict the density and heat of formation of EMs[46], this work is the first of its kind in applying it to directly predict key performance metrics such as detonation velocity and pressure. Our high-throughput workflow, depicted in Figure 1, is a hybrid approach that combines density functional theory (DFT) calculations, thermochemical predictions, graph neural networks, and Bayesian optimization to develop the largest public database of EM-like materials and a surrogate model capable of accurately predicting the detonation performance of molecules containing only C, H, N, and O elements. Further analysis of this rich dataset reveals chemical insights into the factors that drive detonation performance, which can be used to motivate future synthesis studies and new EM development. This work is an advancement over prior approaches to detonation performance modeling due to the breadth of the chemical landscape that the final model encompasses – more than 38k molecules sampled from a screening of over 70B collected from public molecular datasets. By producing a model that achieves close agreement with thermochemical models while remaining generalizable across diverse and previously unexplored regions of chemical space, this study establishes a foundation for accelerated materials discovery and the targeted identification of new energetic compounds.

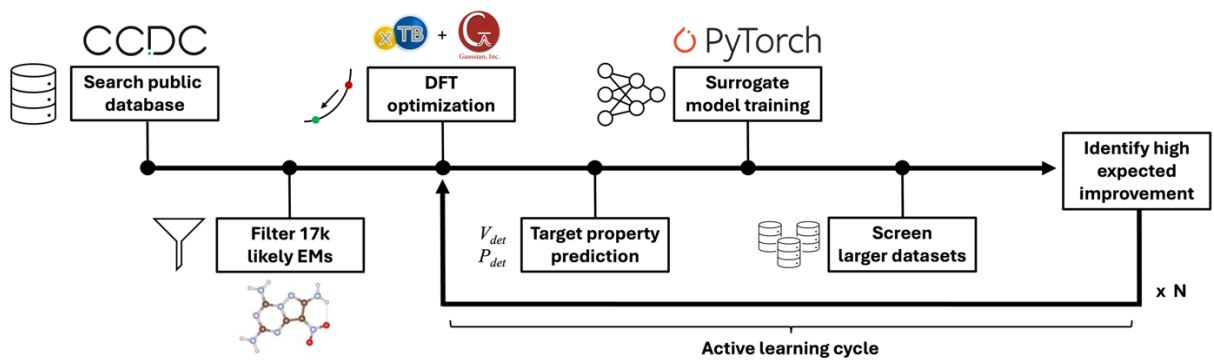

Figure 1 – Schematic of the high-throughput workflow used in this study. An initial surrogate model was trained on a seed dataset of molecules with energetic characteristics and used to screen a larger library of 1.5B molecules sourced from multiple datasets. Expected Improvement,

computed from model predictions and uncertainty estimates, guided the selection of new candidate molecules that balance exploration of underrepresented chemical space and exploitation of high-performing regions. Selected molecules were evaluated with DFT and thermochemical models to obtain detonation performance labels and were then incorporated into the training set. The surrogate model was retrained on the expanded dataset, and this cycle was repeated iteratively until the expected improvement values were sufficiently low, yielding a generalizable model trained on a chemically diverse set of energetic materials.

## 2. Computational Methodology

### 2.1 DFT Calculations

All of the DFT calculations in this work were executed with the Gaussian 16 (vC.01)[47] package. Geometry optimization was conducted at the ωB97X-D/6-311G**[48] level of theory. Our benchmark calculations (Figure *2*) confirm that this level of theory is sufficient to calculate heats of formation ($\Delta H_f$) that are in close agreement with experiment (mean absolute error of 19.3 kJ/mol). Prior to geometry optimization with Gaussian, a conformer search was conducted using the ETKDGv3[49] algorithm as implemented in RDKit.[50] The number of conformers was determined as a function of the number of rotatable bonds in the molecule with a minimum of 50 for rigid molecules and a maximum of 500 for flexible molecules. The resulting conformers were then minimized using the GFN2-xTB[51] tight-binding method through its interface to the Atomic Simulation Environment (ASE).[52] The lowest energy conformer for each molecule was then passed to Gaussian for full DFT relaxation. Job submission and data collection of each DFT calculation was automated using the Python framework *pyiron*.[53]

### 2.2 Evaluation of Detonation Performance

Several packages and algorithms to calculate the detonation properties of EMs are available in the literature.[54–61] In this work, we consider two distinct approaches: the CHEETAH software[54,55] and the Kamlet-Jacobs equations.[56] CHEETAH is a thermochemical code that models the detonation process as an equilibrium transformation of a reactant into a mixture of gaseous and condensed products under the Chapman-Jouguet (CJ) condition.[62,63] CHEETAH's chemical equilibrium formalism minimizes the Gibbs free energy of the detonation products while conserving elemental composition, internal energy, and volume. Thermodynamic data is drawn from an internal database of empirically and theoretically derived equations of state including Becker-Kistiakowsky-Wilson (BKW) for gas phase species and the Murnaghan equations for condensed phase species.[64,65] The CJ equilibrium is determined self-consistently from the intersection of the Rayleigh line and the shock Hugoniot yielding a robust estimate of macroscopic detonation properties.[66] The Kamlet-Jacobs equations introduce a simple algorithm to predict the detonation velocity ($V_{det}$) and detonation pressure ($P_{det}$) of a molecule containing only C, H, N, and O elements. These equations take the following form:

$$V_{det} = 1.01[N^{0.5}M_{avg}^{0.25}Q^{0.25}(1+1.3\rho)] \quad (1)$$

$$P_{det} = 15.58[NM_{avg}^{0.5}Q^{0.5}\rho^2] \quad (2)$$

Where $N$ is the number of moles of gaseous detonation products per gram of EM, $M_{avg}$ is the average molecular mass of the gaseous detonation products in g/mol, $Q$ is the heat release of the detonation reaction (specific enthalpy of explosion) in calories per gram of EM, and $\rho$ is the mass density of the EM in g/cm³. In this work, $Q$ was calculated using the solid-state heat of formation ($\Delta H_f^{solid}$) of the reactant EM. Due to the fact that our DFT calculations are gas phase, we must convert from $\Delta H_f^{gas}$ to $\Delta H_f^{solid}$ by inverting the sublimation reaction as follows:

$$\Delta H_f^{solid} = \Delta H_f^{gas} - \Delta H_{sub} \quad (3)$$

Where $\Delta H_{sub}$ is the heat of sublimation. In this work, $\Delta H_{sub}$ is calculated using the group additive approach of Mathieu[67] as it requires only a molecular graph as input and has been shown to accurately reproduce the experimental $\Delta H_{sub}$ of nitrogen-rich compounds.[20,21,68] This results in predictions of $V_{det}$ in units of km/s and $P_{det}$ in units of kbar.

The heat of explosion ($Q$) is dictated by the set of detonation products accessible to a given explosive. Kamlet and Jacobs originally proposed a simple hierarchical scheme to determine the detonation products for arbitrary CHNO explosives: $N_2$, $H_2O$, and $CO_2$ where oxygen is prioritized to produce $H_2O$ before $CO_2$ and any residual carbon forms graphite.[56] While convenient, this method neglects important detonation products such as CO and $H_2$ and can break down for explosives with an oxygen balance outside of the idealized range. Therefore, in this work we employ the Modified Kistiakowsky-Wilson (MKW) rules[69] to determine the distribution of detonation products that go into calculating $Q$ for the Kamlet-Jacobs equations. CHEETAH, in contrast, uses its own thermochemical equilibrium algorithm to determine detonation products, which we left unchanged.

Density was estimated as follows. For molecules with reported experimental density (sourced from the Cambridge Structural Database[45]), the values were retained without modification. For all other molecules lacking experimental measurements, we used RDKit to compute the molecular volume of the lowest energy MMFF94[70] conformer. These computed volumes were then linearly fit to those with experimental values, and the resulting coefficients were applied to correct the computed volumes. When evaluated on a test subset, this procedure yielded close agreement with experiment (MAE = 0.038 g/cm$^3$ and $R^2$ = 0.87), as shown in Supplemental Information Figure S1.

*2.3 Machine Learning Methods*

We developed a message-passing neural network (MPNN) implemented using the Python package Chemprop[71] as a surrogate model to predict the detonation performance of EMs. In this architecture, atoms and bonds are represented as nodes and edges of a molecular graph. Information is exchanged between bonded atoms through a message-passing procedure analogous to convolution in a conventional neural network. After multiple rounds of message passing, atomic features are pooled into a single molecule-level embedding that is passed to a feed-forward neural network, which predicts the target properties.

Molecular graphs were read as SMILES strings, canonicalized using RDKit, and filtered to remove invalid or duplicate entries. The dataset was partitioned into 80/10/10 training, validation, and test subsets by random selection. This partitioning was repeated ten times to produce unique splits for cross-fold validation. Targets were standardized to the mean and variance of the training set, and the same scaler was applied to the validation and test sets. An inverse-transform layer unscaled the predictions at inference. Molecules were featurized with Chemprop's default *SimpleMoleculeMolGraphFeaturizer* class. A dropout layer with a rate of 20% was applied to the bond message-passing step to prevent overfitting and batch normalization was applied after aggregation to stabilize training. Hyperparameter optimization was managed by the Python package Ray Tune[72] which employs the Asynchronous Successive Halving Algorithm (ASHA)[73] for efficient early-stopping. The grid search conditions are outlined in Supplemental Information Table S1. Each trial was trained for up to 100 epochs with ASHA configured to use a grace period of 20 epochs and a reduction factor of 2. Root mean squared error (RMSE) was chosen as the loss function. An Adam[74] optimizer was used in conjunction with a Noam learning-rate scheduler[75], which scaled the learning rate from an initial value of $1 \times 10^{-4}$ to a final value of $1 \times 10^{-5}$. Model checkpoints were saved for each trial, and the configuration which produced the lowest validation RMSE in each fold was retained for inference. All training was conducted on NVIDIA A100 Tensor Core GPUs, with four GPUs per node used for parallel execution.

The initial training molecules were sourced from the CSD, and their detonation performance targets were calculated using the Kamlet-Jacobs equations and CHEETAH as described in Section 2.2. The database was screened to include only molecules with a valid three-dimensional representation and at least one carbon atom. Additional filters were imposed to retain molecules composed exclusively of C, H, N, and O elements with fewer than sixty non-hydrogen atoms. Finally, molecules lacking N-N, N-O, or O-O bonds were removed, as such species are unlikely to exhibit energetic characteristics. This results in a selection of approximately 17,000 molecules that are herein referred to as the CSD-17k dataset.

As illustrated in Figure 1, the CSD-17k dataset acts as a starting point for exploring the vast chemical space of potential EM candidates, estimated to contain on the order of $10^{60}$ molecules for systems with up to 30 atoms.[76] To sample this broad search space efficiently, we turn to global optimization methods[77] that use acquisition functions to rigorously identify new training data that balance the exploitation of promising regions with the exploration of under-represented space. These acquisition functions are a key component of any AL workflow because they serve to minimize the number of trials required to adequately sample a given search space. In this work, we utilize the expected improvement (EI) acquisition function.[78] For a data point with a mean property value $\mu(x)$ and predictive uncertainty $\sigma(x)$, the EI is defined as:

$$EI(x) = (\mu(x) - f^*)\Phi(Z) + \sigma(x)\phi(Z) \quad (4)$$

$$Z = \frac{\mu(x) - f^*}{\sigma(x)} \quad (5)$$

where $f^*$ is the current best observed value of the target property and $\Phi(Z)$ and $\phi(Z)$ are the cumulative distribution function and probability density function, respectively, of the standard normal distribution. The first term, $(\mu(x) - f^*)\Phi(Z)$, promotes candidates with predicted means exceeding the current best observation. The second term, $\sigma(x)\phi(Z)$, promotes exploration into regions of higher uncertainty. Data points with the largest EI are expected to have the largest gain over the current best-known result, balancing the trade-off between improvement potential and uncertainty.

After each training generation – beginning with the model trained on the initial CSD-17k dataset – the surrogate model was applied to a composite library of more than 70B molecules drawn from GDB[41,42], PubChem[79], ZINC[80], ChEMBL[81] and a number of other patent[82,83] and literature[28,84–87] sources. The library was then screened to include only molecules that passed the previously defined CSD-17k filters and had a synthetic accessibility score (SAScore)[88] below 5 resulting in approximately 1.5B candidates. The SAScore is a heuristic that estimates the difficulty of chemical synthesis by combining molecular complexity, fragment contributions derived from known compounds, and structural features associated with challenging synthesis; lower values indicate more readily synthesizable molecules. Using the model's predictions, the EI value was computed for each molecule, and a batch of approximately 5,000 molecules with the highest EI was selected for incorporation into the training data for the next generation. This procedure was repeated for five generations, terminating when the maximum EI across the full 1.5B molecule library had decreased to just 2 m/s. The resulting collection comprises a diverse dataset of more than 38,000 molecules (including those from CSD-17k), which we refer to as AL-38k throughout this work. We note that, in principle, this process could be repeated for arbitrarily many iterations. However, given the steep reduction in EI values, we expect that further iterations would yield diminishing returns in predictive accuracy.

In addition to the Chemprop models, we also trained a gradient boosting trees (GBT)[89] regression model on a set of topological features computed from molecules in the AL-38k dataset to perform interpretability studies. This model provides a complementary perspective by enabling direct analysis of the structural features that most strongly influence detonation performance. By intentionally excluding DFT-derived descriptors, the analysis is restricted to information of comparable complexity to that available to the graph-based surrogate model. This design choice also highlights which low-cost molecular properties may be most informative for rapid EM screening applications.

The GBT model was implemented using the Python package *scikit-learn* (v1.6.1)[90] and trained on a randomly partitioned 90/10 train-test split. Hyperparameter optimization was performed by further dividing the training set into 20 cross-validation folds, each with a 90/10 train-validate split, and conducting a grid search over the hyperparameters listed in Supplemental Information Table S2. To improve robustness and reduce variance, the optimized GBT model was used as the base estimator in a *scikit-learn BaggingRegressor*[91], forming an ensemble of 10 models trained on bootstrap-resampled versions of the training data (sampling with replacement, with each subset equal in size to the original training set). Predictions from the ensemble were aggregated by taking the mean across all sub-estimators, yielding the final model used for interpretability analysis.

The GBT model was trained on a set of descriptors comprising all two-dimensional features available in the RDKit *rdMolDescriptors* module, along with three additional properties: oxygen balance (%OB), $\Delta H_{sub}$, and density. %OB is known to be highly correlated with detonation performance.[92,93] It can be calculated directly from the stoichiometry of a molecule with the following equation:

$$\%OB = \frac{-1600(2N_C + 0.5N_H - N_O)}{M} \quad (6)$$

where $N_C$, $N_H$, and $N_O$ are the numbers of carbon, hydrogen, and oxygen atoms respectively, and $M$ is the molecular weight. %OB quantifies the extent to which a compound can be oxidized, with values near zero generally considered favorable for maximizing detonation performance.[40] Note that by using Eq. 6, all sources of oxygen in a molecule are counted towards its oxygen balance. This definition is conventional, but differs from an earlier definition by Kamlet and Adolph[94] which excludes oxygen atoms in carbonyl groups. $\Delta H_{sub}$ was calculated using the same group additive approach discussed previously in Section 2.2 and density was evaluated using the fitting approach also outlined in Section 2.2. To prevent redundancy and reduce dimensionality, descriptors exhibiting a Pearson correlation coefficient greater than 0.95 with any other descriptor were filtered out (one descriptor from each highly correlated pair was retained arbitrarily). Descriptors with zero variance were also removed. The remaining descriptors were concatenated to form the input feature vector for the GBT model. A list of the highly correlated descriptors is provided in Supplemental Information Table S3.

Model interpretability was assessed using SHAP (SHapley Additive exPlanations)[95,96], which provides a unified framework for quantifying the contribution of each input feature to an individual prediction. SHAP values were computed using the *TreeExplainer* implementation, which is specifically optimized for ensemble tree models such as the bagged GBT regressor used in this work. For each molecule in the held-out test set, SHAP values were calculated for all input features, producing feature-level attributions that sum to the model's predicted detonation performance. These per-sample attributions were aggregated across the full test set to determine the global importance of each descriptor. Summary statistics, including mean absolute SHAP values and feature impact distributions, were used to identify the descriptors most strongly associated with variations in detonation performance. This analysis enabled direct comparison between the importance of specific topological features and the structure–property relationships inferred from the MPNN surrogate model.

*2.4 Cheminformatics and Structural Analysis*

Molecular structural similarity was quantified using atom-pair fingerprints[97] as implemented in RDKit. In this scheme, each molecule's graph representation is encoded as a fixed-length vector derived from all unique atom–atom pairs, where each pair is characterized by the atom types and the topological distance separating them. Pairwise similarity between molecules was computed using the Tanimoto coefficient, the standard metric for comparing binary or count-based molecular fingerprints.[98,99] The resulting Tanimoto distances were used to construct a quantitative representation of the chemical space spanned by the dataset, with each molecule mapped to a point in the high-dimensional vector space. This representation served as the basis for subsequent structural analyses, including chemical-space visualization, clustering, and the examination of structure–property relationships by projecting molecular properties onto the fingerprint-defined space.

To gain additional insight on the structure-property relationships present in the data, we employed a principal subgraph mining procedure adopted from the work of Kong et al.[100] In this approach, a principal subgraph is defined as a repetitive structural fragment that occurs frequently across many molecules and is not strictly contained within any other more frequent fragment, thereby representing a dominant recurring pattern in the dataset. Starting from all unique atom types as initial fragments, neighboring fragments are iteratively merged to form larger candidate subgraphs. At each iteration, candidate merges are evaluated across the full dataset, and the most frequent merged fragment is added to the vocabulary. The dataset is then updated by replacing occurrences of that fragment with a single token, and the merge-and-update process repeats until a predefined number of principal subgraphs is obtained. This procedure efficiently identifies a vocabulary of frequent molecular substructures without relying on hand-crafted rules or external libraries, providing a basis for downstream structural analysis.

## 3. Results

### 3.1 Detonation Performance

In order to establish a baseline for the performance that can be expected of both CHEETAH and the Kamlet-Jacobs equations, $V_{det}$ predictions from both methods were compared against a dataset of approximately 100 experimental measurements published by Muravyev et al.[101] In addition to the $V_{det}$ measurements, the authors also provide experimental references for $\Delta H_f^{solid}$ which enabled us to assess how sensitive $V_{det}$ is to error in the DFT calculations and subsequent sublimation corrections. Because both methods rely on heat of formation as an input, in Figure 2a,b we compare the performance of B3LYP[102,103] and ωB97X-D[48] DFT functionals in reproducing experimental values for a set of small organic molecules.[104–106] As stated in Section 2.1, the dispersion-corrected ωB97X-D functional exhibits markedly better performance than B3LYP and as such is used for all DFT calculations. It is important to note that while the choice of ωB97X-D was based on superior performance in calculating $\Delta H_f^{gas}$, the inputs passed to both the Kamlet-Jacobs equations and CHEETAH are converted from $\Delta H_f^{gas}$ to $\Delta H_f^{solid}$ using the procedure outlined in Section 2.2. A direct comparison of the $\Delta H_f^{solid}$ from experiment and corrected DFT calculations is provided in Supplemental Information Figure S2.

Subplots (c) and (d) of Figure 2 illustrate the performance of the Kamlet-Jacobs equations (KJ) and CHEETAH respectively when using DFT-derived $\Delta H_f^{solid}$. Both methods achieve a high coefficient of determination ($R^2$) and low mean absolute error (MAE) relative to experiment. However, CHEETAH achieves a notably smaller MAE (189 m/s vs. 251 m/s), which indicates that the proprietary method more accurately captures the experimental conditions than the more general Kamlet-Jacobs equations. A similar figure illustrating the performance of both methods when using experimental $\Delta H_f^{solid}$ as input is presented in Supplemental Information Figure S3. No improvement in predictive accuracy is observed when using the experimental $\Delta H_f^{solid}$ over the DFT-derived $\Delta H_f^{solid}$. This supports the idea that DFT, in conjunction with a thermochemical code like CHEETAH, can be used as an accurate proxy to investigate the theoretical detonation performance of EMs in a high-throughput framework.

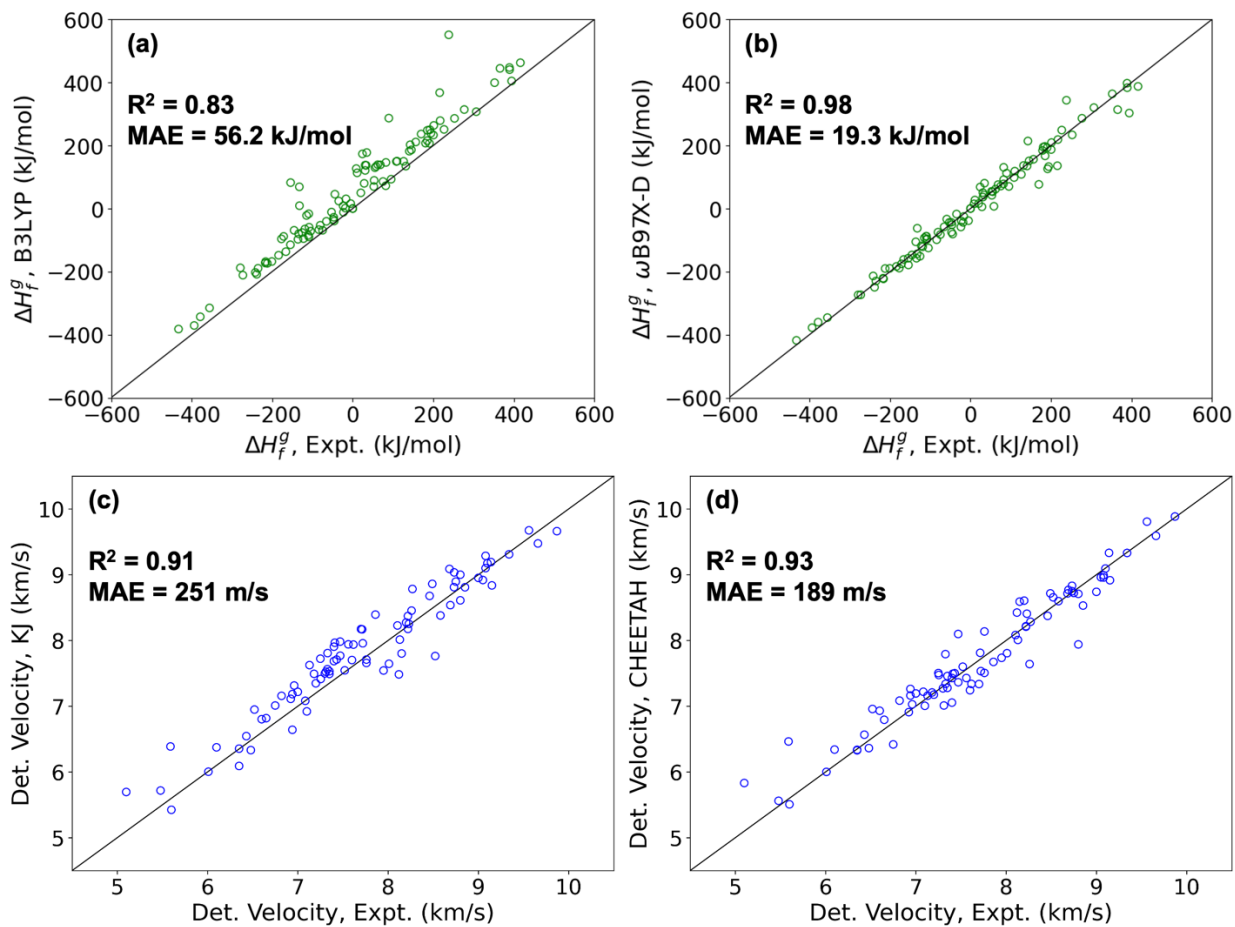

Figure 2 – Parity plots comparing the performance of different methods to predict experimentally measured $\Delta H_f^{gas}$ and $V_{det}$. Subplots (a) and (b) compare the performance of B3LYP and ωB97X-D functionals in reproducing $\Delta H_f^{gas}$ for a set of small organic molecules. Subplots (c) and (d) compare the performance of the Kamlet-Jacobs equations and CHEETAH in predicting experimentally measured $V_{det}$ using DFT-derived $\Delta H_f^{solid}$ and experimental density as inputs.

After verifying our approach, high-throughput DFT calculations to optimize geometry and derive $\Delta H_f^{solid}$ of the molecules in the CSD-17k database were conducted using the *pyiron*[53] Python framework. Both CHEETAH and the Kamlet-Jacobs equations were applied to each molecule to obtain $V_{det}$ and $P_{det}$. The property distributions predicted by both methods can be seen in Figure 3. Subplots (a) and (b) are joint plots illustrating the relationship between $V_{det}$ and $P_{det}$. Both methods show the same exponential trend between the two properties, but the values obtained from CHEETAH exhibit a more disperse distribution. Furthermore, the vast majority of molecules have a $V_{det}$ less than 6 km/s (91% for Kamlet-Jacobs and 84% for CHEETAH), highlighting just how rare performant EMs are and the necessity to screen a broad chemical landscape for adequate candidates. In subplots (c) and (d), the Kamlet-Jacobs and CHEETAH predictions of both properties are superimposed to illustrate systematic differences between the two methods. In Figure 3c it is clear that CHEETAH tends to predict higher average $V_{det}$ relative to the Kamlet-Jacobs equations. This effect is less substantial, though still evident, in Figure 3d for $P_{det}$.

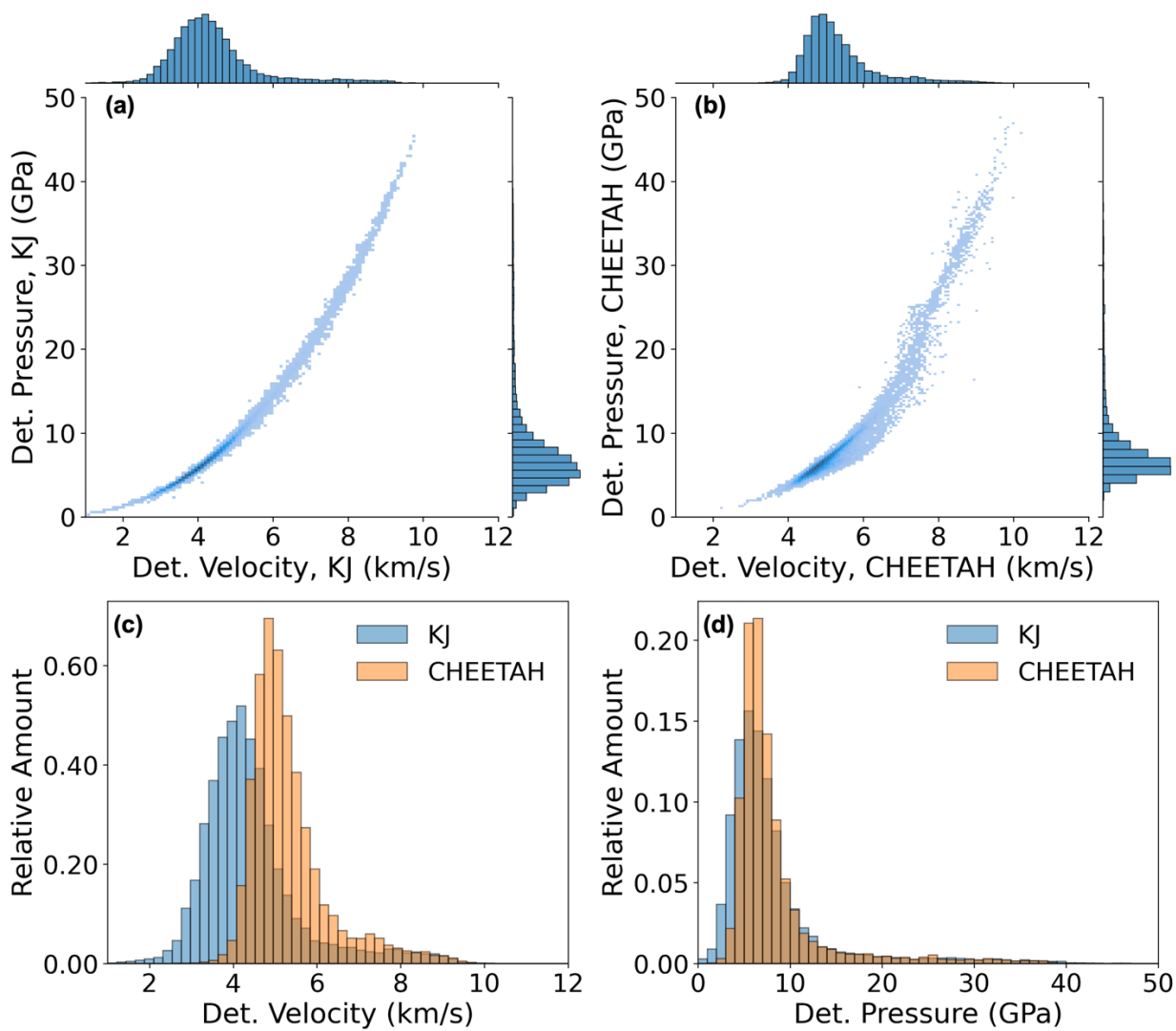

Figure 3 – Distributions of $V_{det}$ and $P_{det}$ as predicted by CHEETAH and the Kamlet-Jacobs equations for molecules in the CSD-17k dataset. Subplots (a) and (b) illustrate the relationship between $V_{det}$ and $P_{det}$ as predicted by the Kamlet-Jacobs equations and CHEETAH respectively. Subplots (c) and (d) illustrate systematic differences in each method's prediction of $V_{det}$ and $P_{det}$ respectively.

To investigate the underlying cause of this systematic difference, we analyzed the distribution of detonation products predicted by both methods. For consistency, the analysis was conducted at the Chapman-Jouguet[62,63] (CJ) point which is implicit in the Kamlet-Jacobs equations and the MKW ruleset used to determine product distributions. CHEETAH, being a comprehensive thermochemical code with additional applications outside of detonation performance prediction, considers a larger set of possible product species than the MKW rules do. This larger product list means that CHEETAH can predict products that MKW does not and for overlapping products there may be differences in how the available C, H, N, and O atoms are prioritized to produce them. The predicted consumption of each element by the nine most prevalent detonation products of molecules in the CSD-17k dataset is illustrated in Figure 4.

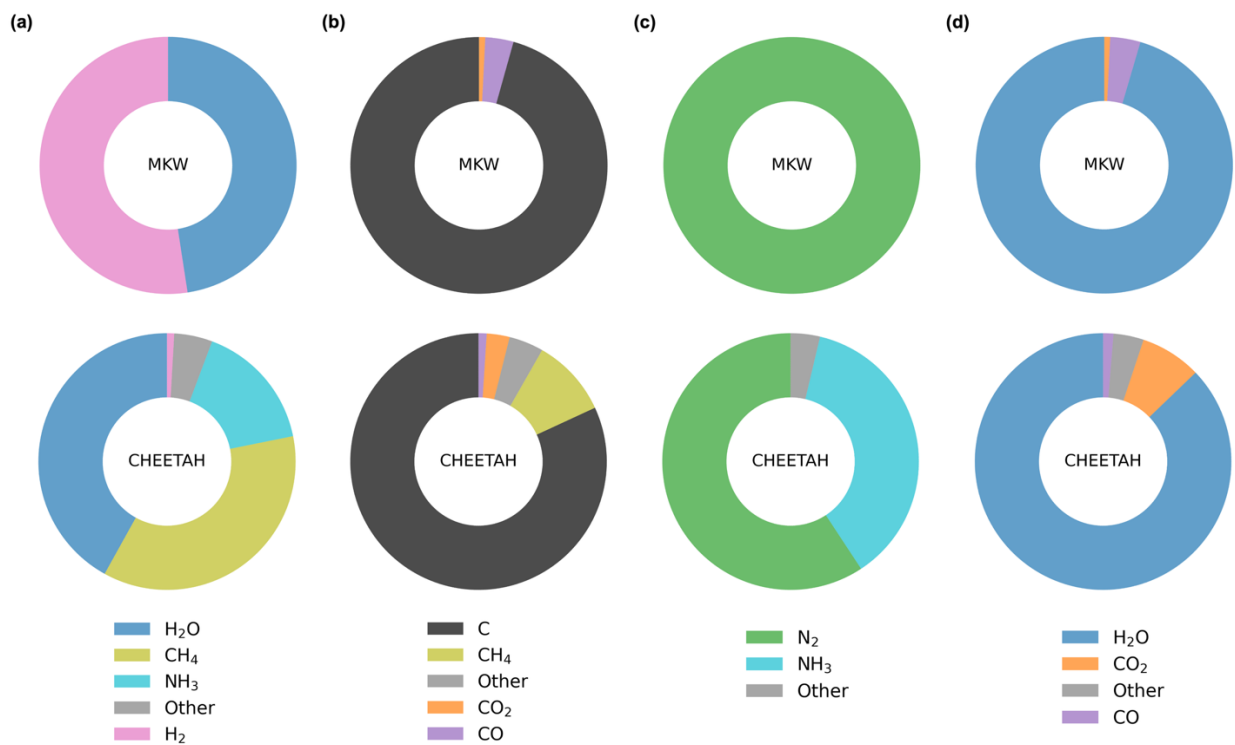

Figure 4 – Element-wise breakdown of the decomposition products predicted by CHEETAH and the MKW rules. Each pie chart illustrates the percentage of the total available amount of an element that gets consumed by each detonation product. The percentages are averaged over all molecules in the CSD-17k dataset. Each subplot corresponds to a different element: (a) hydrogen, (b) carbon, (c) nitrogen, (d) oxygen.

For the case of H (Figure 4a), the MKW formalism allows for just two H-consuming products, $H_2O$ and $H_2$. CHEETAH, however, also allows H to be consumed by $CH_4$ and $NH_3$. It should be noted that CHEETAH in fact predicts many other H-containing species, but their total contribution is quite small and thus gets categorized as a singular "Other" class for clarity. There is a stark difference between these two as MKW predicts more than half (52.4%) of all H going to $H_2$ whereas the $H_2$ production predicted by CHEETAH is almost negligible (0.9%). Both methods predict similar amounts of H going to $H_2O$ (47.6% and 41.9% for MKW and CHEETAH respectively), while the remainder in CHEETAH is primarily consumed by $CH_4$ and $NH_3$ (both products which are not considered by the MKW rules). The abundance of $H_2$, which has $\Delta H_f = 0$ by definition, predicted by the MKW rules highlights how H under the MKW formalism is being consumed by less energetically favorable products compared to CHEETAH. This results in a less negative heat of explosion and thus a reduced $V_{det}$ relative to CHEETAH.

Another significant difference is the case of N (Figure 4c). Under the MKW rules, all N goes to $N_2$. CHEETAH predicts that the majority (59.2%) of N goes to $N_2$, but a significant amount (37.0%) is used to produce $NH_3$. Here again, the MKW rules are observed to predict a large amount of product with $\Delta H_f = 0$ whereas CHEETAH tends to predict a distribution of products with a more negative $\Delta H_f$ on average leading to systematic differences in $V_{det}$.

The consumption of C (Figure 4b) and O (Figure 4d) are largely similar between the two methods indicating that the observed offset in $V_{det}$ is primarily driven by the treatment of H- and N-containing detonation products. A more detailed version of the data in Figure 4 is presented in Supplemental Information Figure S4 as a series of histograms to avoid information loss from averaging. Alternatively, Supplemental Information Figure S5 compares the distribution of the number of moles of each detonation product predicted by both methods.

*3.2 Molecular Structure Analysis*

Figure 5 illustrates the results of molecular fingerprint-based structural analysis (Section 2.4) applied to the CSD-17k dataset with detonation performance predictions taken from CHEETAH. Atom pair fingerprints[97] are chosen as the specific fingerprinting algorithm and t-SNE[107] with a Tanimoto distance kernel is used to reduce their dimensionality for visualization. Additional visualizations of the same dataset using alternative fingerprinting algorithms and dimensionality reduction techniques can be found in the Supplemental Information Figures S6-9. In Figure 5a, each molecule's detonation velocity is depicted with a colormap ranging from blue (low $V_{det}$) to red (high $V_{det}$). It is clear from this subplot that the highest performers are spatially clustered and distinct from the lowest performers. Furthermore, the most extreme regions are separated from one another by molecules with intermediate performance. This clustering of extremes and the gradient between them indicates a strong correlation between structure and performance that can be leveraged to inform future synthesis and predictive model development. Another interesting feature in the data is the separation of the highest performing molecules into two distinct spatial regions. This suggests that different structural motifs can be equally performant, which affords more flexibility in the design constraints for new EM development. In comparing the members of encircled regions *b* and *c* (Figure 5b,c) it is clear that region *b* corresponds to molecules with a linear backbone that are rich in trinitromethyl groups, while the molecules in region *c* consist of linked aromatic rings. The consistent feature between the two is the heavy presence of $NO_2$ groups, which are known to contribute to detonation performance.[108] In low-performance region *d* (Figure 5d), the molecules are unadorned by $NO_2$ groups or any other explosophore and instead contain carbonyls and heteroatoms in the backbone which results in an underutilization of N and O atoms that could be more effectively used to synthesize energetic functionalities. These observations are consistent with expectations and reinforce $NO_2$ density as a key driver of detonation performance.

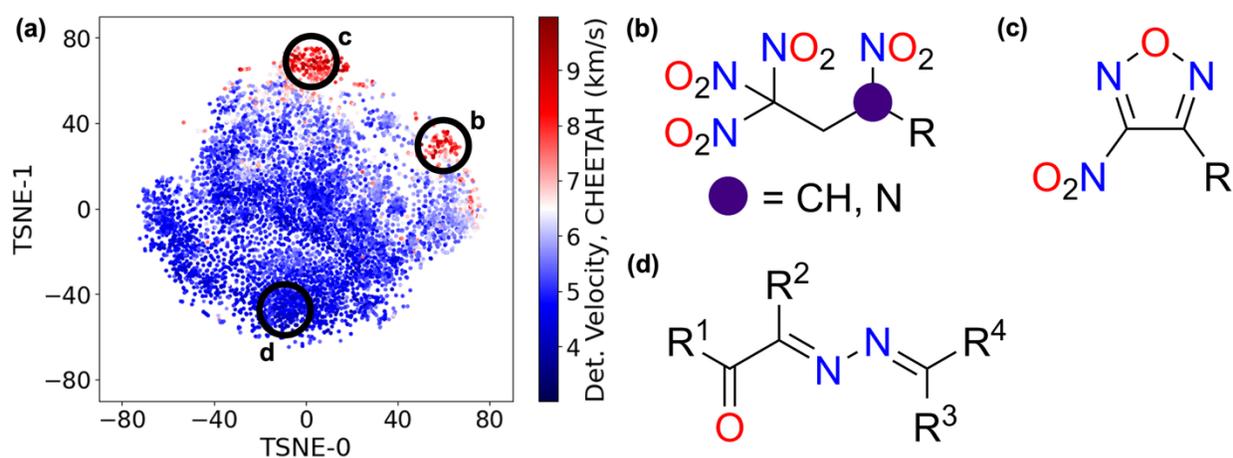

Figure 5 – (a) Projection of the atom pair fingerprints of the CSD-17k molecules into 2D space using t-SNE for dimensionality reduction. The color bar illustrates the $V_{det}$ of each molecule with blue points indicating a low value and red points indicating a high value. The molecular motifs in subplots (b), (c), and (d) are derived from the nearest neighbors of the centroid of their corresponding encircled region in subplot (a).

In addition to the analysis of molecular fingerprints presented above, principal subgraphs of the highest and lowest performers are identified using the methodology of Kong et al.[100] (Section 2.4). Substructures identified as being most significant to molecules in the 90[th] percentile of $V_{det}$ include cubane-like cages, especially those connected to $NO_2$ groups. Substructures from molecules in the 10[th] percentile of $V_{det}$ contain no $NO_2$ groups and are primarily composed of saturated C-backbones or aromatic rings. Visualizations of the principal subgraphs in each performance group are presented in Supplemental Information Figure S10.

*3.3 Active Learning*

As discussed previously, the large chemical design space of potential EM candidates necessitates the development of a generalizable model for effective and reliable high-throughput screening. To achieve this, we employed an active learning cycle (Section 2.3) that, through a Bayesian optimization framework, selects a minimal set of new training samples based on their expected improvement (EI) scores. This approach prioritizes molecules for which the model is either highly uncertain or predicts strong performance, thereby expanding the training set in a directed way that enhances both chemical diversity and exposure to highly performant EMs. Figure 6 summarizes the results of this iterative process.

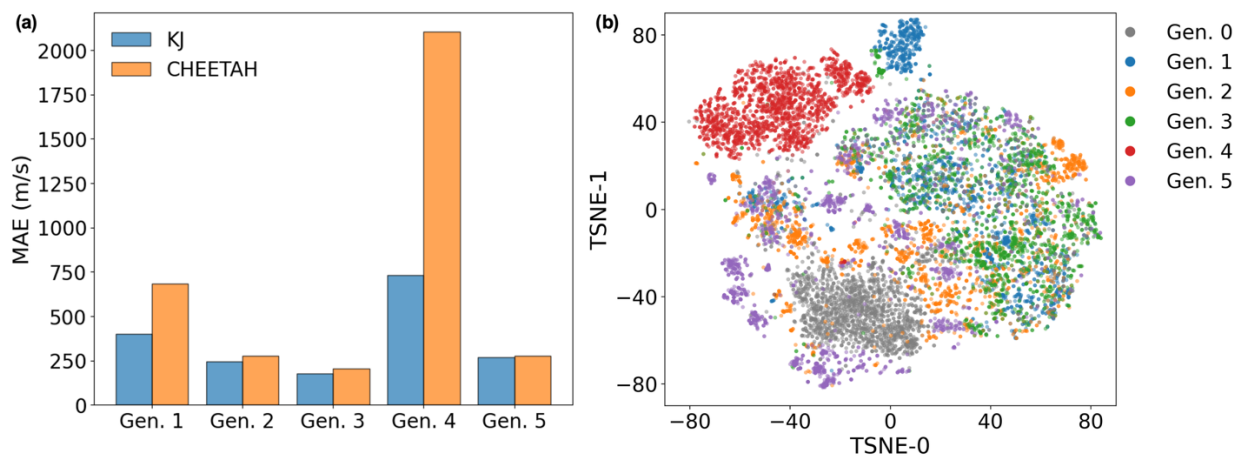

Figure 6 - (a) Performance of each active learning model trained on molecules from Gen. 0 (CSD-17k) through Gen. *X*-1 in predicting unseen molecules from Gen. *X*. (b) Chemical space spanned by each active learning generation. Each point is a t-SNE projection, calculated from a Tanimoto distance matrix, of a molecule's atom pair fingerprint. Points are color coded by their generation number with the CSD-17k dataset shown in gray.

Figure 6a depicts the performance of each active learning model on previously unseen (out-of-sample) molecules compared to values calculated using the Kamlet-Jacobs equations and CHEETAH. More specifically, each bar reports the MAE of a model trained on molecules from Gen. 0 (CSD-17k) through Gen. *X-1,* where *X* corresponds to the label on the *x*-axis. This is a convenient way to track how well each iteration of the model generalizes to newly selected molecules with large EI values. Figure 6b illustrates the chemical space spanned by all molecules in the AL-38k dataset, color coded by their generation number. An alternative view, in which each generation is plotted sequentially, can be found in Supplemental Information Figure S11. As in Figure 5a, each point corresponds to the t-SNE projection of a molecule's atom pair fingerprint, ensuring that points which are physically clustered are also chemically similar. From inspection, the molecules in Gen. 0 (colored gray) form a distinct subset that have minimal overlap with molecules from subsequent generations. This manifests as a significant drop in predictive performance when the Gen. 0 model is applied to molecules from Gen. 1 (colored blue) – the first batch of molecules with high EI selected from the composite dataset of more than 1.5 billion. The model achieves a MAE of 112 m/s relative to the Kamlet-Jacobs equations on the in-sample CSD-17k test set (Supplemental Information Figure S12); however, when extrapolated to the previously unseen chemistry of Gen. 1 molecules, that error increases substantially to 399 m/s. The CHEETAH predictions are similarly affected – MAE of 173 m/s on the Gen. 0 in-fold test set rising to 683 m/s when applied to Gen. 1 molecules. Additional analysis of the Gen. 1 molecules was conducted to identify which structural features contributed most to the model's reduced accuracy. We identified that Gen. 1 molecules were more likely to contain at least one hydroxyl (-OH) group relative to Gen. 0 (36% v. 26%) and significantly more likely to contain multiple -OH groups (15% v. 7%). This compositional difference is consistent with the clear separation between Gen 0. and Gen. 1 clusters observed in Figure 6b. A detailed breakdown of this analysis is provided in Supplemental Information Figure S13. As additional chemistry was incorporated through Gens. 2 and 3 (each one expanding the model's training set with new batches of molecules with high EI), the model's predictive performance on out-of-sample molecules improved steadily, reaching a minimum MAE of 177 m/s relative to the Kamlet-Jacobs equations and 206 m/s relative to CHEETAH calculations. As evidenced by Figure 6b, the chemical space spanned by Gen. 4 molecules (colored red) is largely disjoint from that of the prior generations. This divergence is reflected in the pronounced spike in prediction error shown in Figure 6a, where the MAE relative to the Kamlet-Jacobs equations increases to 730 m/s and the MAE relative to CHEETAH calculations rises to 2,105 m/s. Structural analysis of Gen. 4 molecules, presented in Supplemental Information Figure S14, revealed high concentrations of previously unseen 1,3-dioxolane rings which likely contributed to the significant drop in predictive performance. The distinct composition of Gen. 4 molecules highlights the ability of the active learning framework to sample both exploitative and explorative regions of the chemical landscape – each of which is crucial in developing a generalizable and effective surrogate model. After incorporating the unique chemistry of Gen. 4 molecules into the training set, the model rapidly recovered its predictive performance – achieving a MAE of 267 m/s relative to the Kamlet-Jacobs equations and 275 m/s relative to CHEETAH calculations. This high level of out-of-sample accuracy is consistent with the distribution of molecules in Gen. 5 (colored purple), which are observed to primarily "fill in the gaps" between preceding generations rather than expanding into new regions of chemical space. Notably, this final model exhibits the smallest gap in performance between Kamlet-Jacobs and CHEETAH targets (a difference of only 8 m/s), indicating that CHEETAH's complex objective function can be learned as effectively as Kamlet-Jacobs' when trained on sufficiently diverse data. Additional plots that show the distribution of ground truth detonation performance values for all AL-38k molecules as calculated by both methods can be found in Supplemental Information Figure S15.

While the active learning loop clearly improved the model's generalizability relative to its initial state, it is important to note that the cycle could, in principle, continue indefinitely to incorporate increasingly exotic regions of chemical space. In practice, however, the rapid decline in EI across generations indicates that additional rounds of training would yield diminishing returns. As a practical application of the model in its intended role, we screened the composite 1.5B molecules dataset using the finalized Gen. 5 weights. More than 1M molecules were identified for which the predictions of both Kamlet-Jacobs and CHEETAH targets agreed would have a $V_{det}$ above 6 km/s. Approximately 10k of these molecules had a mutually agreed upon $V_{det}$ of over 7.5 km/s, making them excellent candidates for future investigation.

*3.4 Feature Importance*

To complement the MPNN surrogate model with a more interpretable perspective, we evaluated the feature importances learned by the GBT model (Section 2.3) to determine which molecular descriptors contribute most to detonation performance. Figure 7 illustrates the SHAP scores of the six most important descriptors in the GBT model's prediction of $V_{det}$ relative to CHEETAH ground truth. SHAP scores for $P_{det}$, along with parity plots demonstrating the GBT model's predictive accuracy, are presented in Supplemental Information Figures S16-20. Extended descriptions of the most important descriptors are provided in Supplemental Information Table S4.

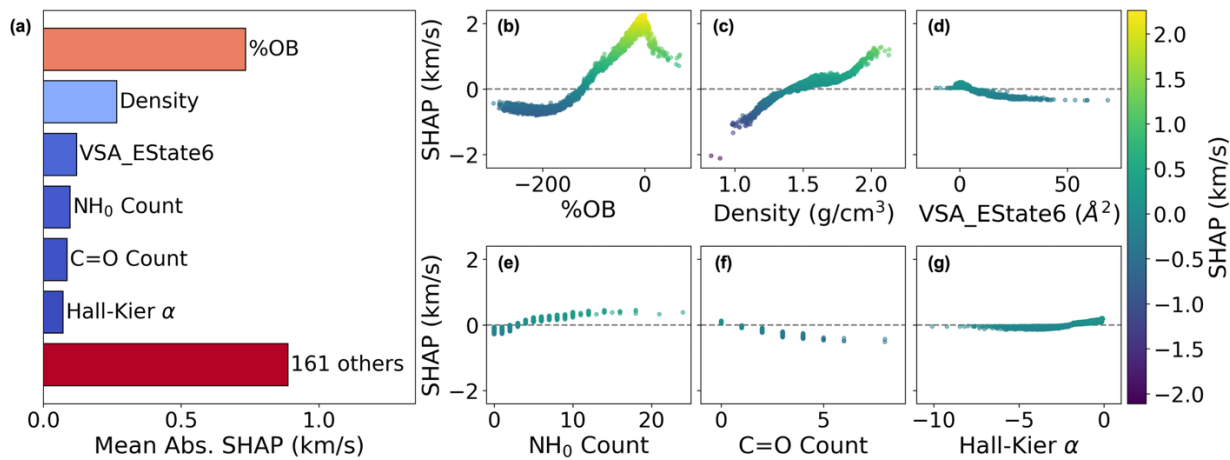

Figure 7 – SHAP importance scores of the top descriptors in predicting $V_{det}$ as calculated by CHEETAH. (a) Mean absolute SHAP score for the top six most important descriptors alongside the sum of all other descriptors. Error bars are not illustrated because the standard deviation of the mean absolute SHAP score is ≤ 13 m/s for each individual descriptor. (b-g) Scatter plots of SHAP score v. descriptor value for each molecule in the GBT model's test set. Each subplot corresponds to a different descriptor: %OB, Density, VSA_EState6, $NH_0$ Count, C=O Count, and Hall-Kier $\alpha$ respectively. The colormap scales linearly with the Y-axis to emphasize data points with the most extreme SHAP scores. Positive SHAP values are associated with a positive contribution to predicted $V_{det}$ while negative SHAP values are associated with a negative contribution to predicted $V_{det}$.

The resulting feature importance ranking reveals a clear hierarchy among the descriptors, with oxygen balance (%OB) emerging as the dominant contributor. Oxygen balance exhibits more than twice the mean absolute SHAP score of any other individual descriptor under consideration (734 +/- 13 m/s). Given that the energy released by a chemical explosive is primarily the result of the oxidation of carbon and hydrogen, it is reasonable to expect (and indeed already known) that this property would have a strong influence on downstream performance metrics like $V_{det}$ and $P_{det}$.[109] Subplot Figure 7b shows how the SHAP score changes with the value of the %OB descriptor measured for each molecule in the test set of the GBT model. A clear peak in the SHAP score emerges as %OB approaches zero. However, as %OB becomes positive, there is a reduction in the importance of the descriptor highlighting a tendency for performant EMs to have a slightly negative oxygen balance. Furthermore, for molecules with an oxygen balance less than approximately -125, the SHAP score becomes negative, indicating that excessive oxygen deficiency is associated with a reduction in predicted $V_{det}$.

Density is the second most important feature (266 +/- 6 m/s), consistent with its well-established correlation with detonation performance.[108,110] In subplot Figure 7c, the SHAP score is observed to increase monotonically with density. Below a density of approximately 1.4 g/cm$^3$ the SHAP score becomes negative suggesting that molecules with a density below this threshold are unlikely to be performant EMs while those with a density near 2.0 g/cm$^3$ are strongly favored.

The sixth component of the VSA_EState family of descriptors ranks third (120 +/- 9 m/s). These descriptors capture how much of a molecule's van-der Waals surface area is associated with atoms in specific electronic environments, effectively blending electrotopological character with steric accessibility. This is chemically meaningful in the context of EMs because the electronic and steric character of reactive sites dictates decomposition pathways, bond-breaking energetics, and ultimately the rate and extent of gaseous product formation during detonation.[111] The fact that an individual component of VSA_EState has a high importance suggests that detonation performance is not only sensitive to global compositional features like oxygen balance, but also to the distribution of local chemical environments within the molecule – akin to substituent effects and patterns of electron density which are more commonly considered aspects of EM design. While a clear physical interpretation of the structural features associated with any individual component of this family of descriptors is difficult to ascertain, the trend observed in Figure 7d indicates that an increasing contribution to the molecular surface area from the sixth component in particular is associated with a decrease in the SHAP score. This suggests that the abundance of a specific chemical environment, likely one common to the low-$V_{det}$ molecules illustrated in Figure 5d, contributes negatively to the model's prediction of detonation performance.

Two purely functional group-based descriptors, NH$_0$ Count (the number of nitrogen atoms without any bonds to hydrogen) and C=O Count (the number of carbonyl groups), have comparable influence on $V_{det}$ predictions (97 +/- 5 m/s and 86 +/- 3 m/s respectively). However, Figure 7e,f indicate that they affect the model in opposite directions. The interpretation of C=O Count is straightforward – as the number of carbonyl groups in a molecule grows, the SHAP score becomes increasingly negative. This complements the analysis of Figure 5d which shows that carbonyl groups are a common structural component of the lowest-$V_{det}$ molecules in the CSD-17k dataset. Furthermore, this interpretation is aligned with the work of Kamlet and Adolph[94] who consider oxygen atoms in carbonyl groups to be "dead weight" in the sense that they do not contribute to the formation of energetically favorable detonation products. Analysis of the $NH_0$ Count descriptor is complicated by the fact that the SMARTS pattern used to identify this substructure in a molecular graph ([NH0, nH0]) matches N atoms in explosophores such as $NO_2$ groups as well as N atoms in stable aromatic rings or branched aliphatic networks. While the intention of this descriptor as written in the RDKit documentation is to identify tertiary amines, in practice, the descriptor captures a much broader range of chemical environments. Nevertheless, the positive association between $NH_0$ Count and SHAP score suggests that, on average, the increased presence of $NH_0$ environments is correlated with an increase in predicted $V_{det}$. This finding is consistent with the high concentration of $NO_2$ groups observed in the highest-$V_{det}$ molecules presented in Figure 5b,c. However, the importance is likely muted by the fact that nitrogen atoms in $NC_3$ or C=N-C type environments also match the pattern but are common to molecules with low detonation performance. It should be noted that $NO_2$ Count is also a descriptor which the model is trained on, but its SMARTS pattern ([$([NX3](=O)=O), $([NX3+](=O)[O-])][!#8]) does not match $O-NO_2$ linkages, which are a key component of powerful EMs like pentaerythritol tetranitrate (PETN).[112,113] This likely reduces the SHAP score of the $NO_2$ Count descriptor relative to the more flexible $NH_0$ Count descriptor (which does count the nitrogen in the $O-NO_2$ linkage) causing it not to appear in the top six descriptors despite having a clear influence on detonation performance in reality.

Despite having a relatively high mean absolute SHAP score (71 +/- 8 m/s), interpretation of the Hall-Kier $\alpha$ descriptor is made more challenging by the fact that very little change in the SHAP score is observed as a function of the descriptor value (Figure 7g). In general, the descriptor is interpreted as an atom-type correction index computed as the sum of per-atom contributions derived from covalent radii relative to a $sp^3$ carbon reference.[114] In CHNO molecules, this descriptor serves as a proxy for heteroatom substitution patterns and their associated steric effects. However, in the absence of a clear relationship between SHAP score and descriptor value, we conclude that while heteroatom substitution is correlated with predicted $V_{det}$, the directionality of this relationship remains unclear, and no specific design constraints can be inferred.

A previous study by Davis et al. that also sought to rank the feature importance of a descriptor-based model of $V_{det}$ found that the dominant descriptors were $M \log P$ (the water/octane partition ratio), the average molecular weight of the gaseous detonation products, and MolDensity (a measure of density derived from the volume of a molecule in the gas phase).[20] Notably, the study did not include oxygen balance explicitly as a descriptor. Despite the difference in chosen descriptors, their model also achieved good predictive performance of $V_{det}$ relative to the Kamlet-Jacobs equations (RMSE of 110 m/s and $R^2$ of 0.99 on their test set). These results indicate that detonation performance can be learned from multiple, complementary descriptor sets, and that well-constructed models are relatively flexible to the specific choice of molecular descriptors, provided those descriptors capture the relevant underlying chemical phenomena.

Taken together, these insights suggest that high-performance EM candidates tend to exhibit readily oxidizable compositions (slightly negative oxygen balance), high density, and minimal incorporation of carbonyl groups, which are associated with low-performance structural motifs. Descriptors such as VSA_EState and Hall-Kier $\alpha$ are more challenging to derive actionable design constraints from but provide nuanced guidance that suggests local electronic structure effects and specific substituent patterns, not just global formulation metrics, can impact overall detonation performance.

## 4. Conclusions

In this work, we developed an active learning framework that enables generalizable prediction of detonation performance across a broad region of CHNO chemical space. It approaches the accuracy of CHEETAH and the Kamlet-Jacobs equations but requires only the molecular graph as input - making it orders of magnitude faster than either method or any other method that requires high-fidelity estimates of material properties as input. By iteratively retraining a message-passing neural network surrogate model on molecules selected through an expected improvement acquisition strategy, the model progressively expanded its domain of competence while requiring only a modest number of additional training samples per generation. This approach yielded a final training set exceeding 38,000 molecules - more than double the size of the initial dataset and substantially larger and more diverse than those used in prior modeling efforts. As a result, the surrogate model demonstrated markedly improved extrapolative capability relative to the initial baseline and maintained high predictive accuracy even when evaluated on previously unexplored classes of molecules. These results underscore the importance of training on chemically diverse datasets when developing surrogate models intended for broad applicability, highlighting how targeted expansion of chemical space via iterative training can enhance generalization beyond what is achievable through static data collection alone. Furthermore, evaluating the similarity between the chemical space of prospective screening targets and that of the surrogate model's training set should be considered best practice to ensure an effective deployment of such models.

Beyond predictive performance, this work provides insight into the underlying structure–property relationships that govern the theoretical performance of energetic molecules and materials. A complementary GBT model trained on topological descriptors derived from the active learning dataset enabled SHAP-based feature importance analysis, confirming that oxygen balance is the dominant driver of detonation performance, with additional contributions from density, local electronic environments, and a negative influence from the presence of oxygen-consuming functional groups such as carbonyls. Unsupervised cheminformatics analysis of molecular fingerprints reveals how EMs with similar detonation performance profiles tend to cluster together in chemical space and highlights how high performance can be associated with multiple distinct structural motifs. Furthermore, analysis of the decomposition products predicted by different formalisms underscores how sensitive predicted detonation performance can be to assumptions in the underlying chemical equilibrium model. These findings reinforce established chemical intuition while also highlighting more subtle, descriptor-level signals that emerge when learning from a chemically diverse dataset.

Together, these results demonstrate that active learning provides a powerful, data-efficient route for constructing fast surrogate models capable of navigating the chemical landscape relevant to energetic materials. The surrogate developed here is not only more generalizable than models trained on static datasets but also paired with interpretable analyses that reveal actionable design principles for future energetic materials development. Looking forward, this framework naturally integrates with molecular generative models: the active-learning-trained surrogate offers a fast, reliable scoring function that can be embedded directly into optimization loops, reinforcement-learning agents, or diffusion-based generators tasked with proposing novel CHNO structures. In this setting, the surrogate serves as an evaluative engine that can rapidly guide generative models toward chemically feasible, high-performing candidates. The success of this approach suggests a clear path toward a closed-loop discovery workflow in which generative design, surrogate-model evaluation, and targeted high-fidelity computations operate in coordination to accelerate the identification of next-generation energetic materials.

**Acknowledgements**

Research presented in this article was supported by the Laboratory Directed Research and Development program of Los Alamos National Laboratory under project number 20250006DR. This research used resources provided by the Los Alamos National Laboratory Institutional Computing Program, which is supported by the U.S. Department of Energy National Nuclear Security Administration under contract No. 89233218C-NA000001. Additional funding was provided by the NNSA Minority Serving Institution Partnership (MSIPP). The authors thank Frank Marrs and Jack Davis for helpful discussions. This work has been authorized for unlimited release under LA-UR-26-22356.

**Author Contributions**

IM, WKK, and CJS conceived the project. RSU, MJC, IM, and JNS developed the high-throughput workflows to run density functional theory calculations and detonation performance calculations. RSU created the AL-38k dataset and performed molecular structure analysis and surrogate model predictive performance analysis. WKK curated the training data and trained the active learning surrogate models. MCD performed feature importance analysis. RSU wrote the manuscript with input from IM, WKK, MJC, AHS, and CJS. All authors revised and approved of the final manuscript.

**Data Availability**

A subset of the AL-38k dataset can be made available upon reasonable request.

**Keywords**

Supplemental Information for the Manuscript:

# Active Learning for Generalizable Detonation Performance Prediction of Energetic Materials

R. Seaton Ullberg[1,*], Megan C. Davis[1], Jeremy N. Schroeder[1,2], Andrew H. Salij[1], M. J. Cawkwell[1], Christopher J. Snyder[3], Wilton J. M. Kort-Kamp[1], Ivana Matanovic[1,*]

[1]Theoretical Division, Los Alamos National Laboratory, Los Alamos, NM 87545, USA

[2]Department of Mechanical Engineering, Texas Tech University, Lubbock, TX 79409, USA

[3]Weapon Stockpile Modernization Division, Los Alamos National Laboratory, Los Alamos, NM 87545, USA

*Corresponding authors: sullberg@lanl.gov, ivana@lanl.gov

**Table of Contents**

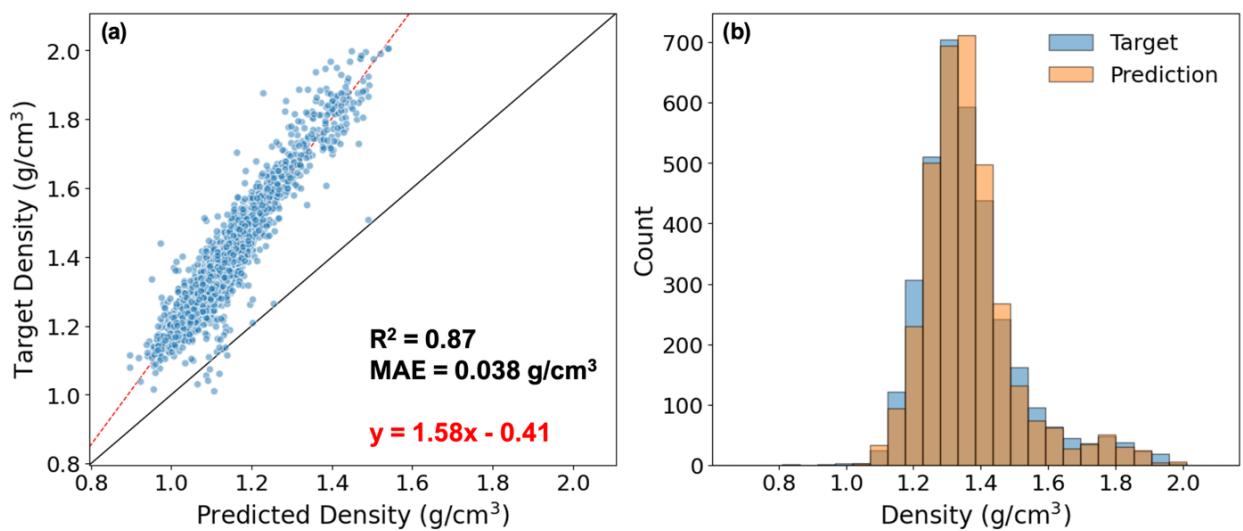

Figure S1 – Comparison of fitted density and experimental ground truth for a test subset of the CSD-17k molecules. (a) Parity plot illustrating the difference between predicted and actual density. The dashed red line represents the line of best fit which is used as a linear transform to correct the molecular volume approach of RDKit. $R^2$ and MAE are measured relative to the line of best fit. (b) Histogram showing close agreement in the distributions of predicted and actual density.

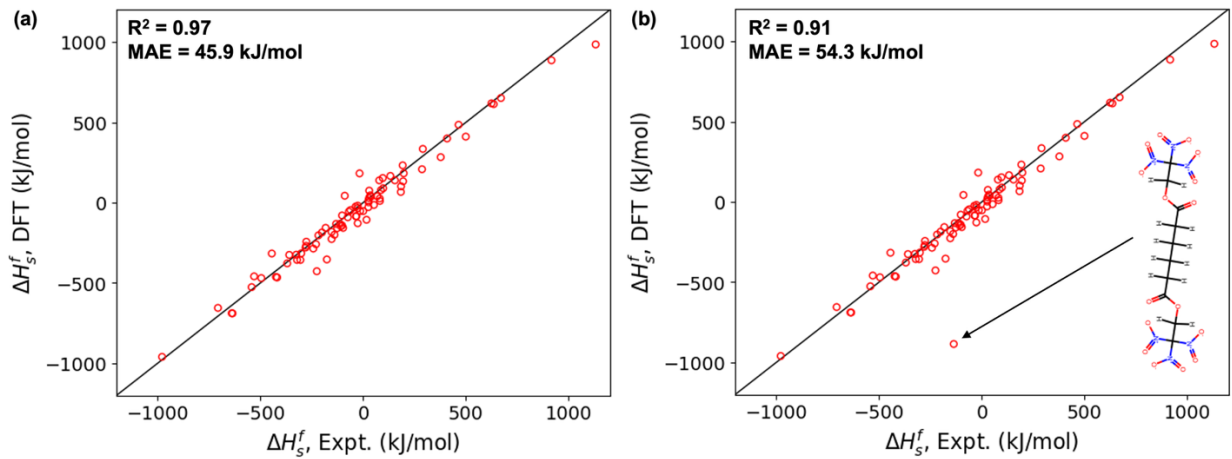

Figure S2 – Performance of DFT (ωB97X-D/6-311G** level of theory) in predicting the solid-state heat of formation ($\Delta H_s^f$) for a set of energetic molecules with experimental reference values from Muravyev et al.[1]. The translation from gas phase (the ab-initio state) to solid phase (the experimental state) is handled by incorporating the heat of sublimation ($\Delta H_{sub}$) as predicted by the group additive approach of Mathieu[2]. Subplot (a) illustrates the performance when a lone outlier is excluded, and subplot (b) shows the offending molecule and the resulting degradation in performance.

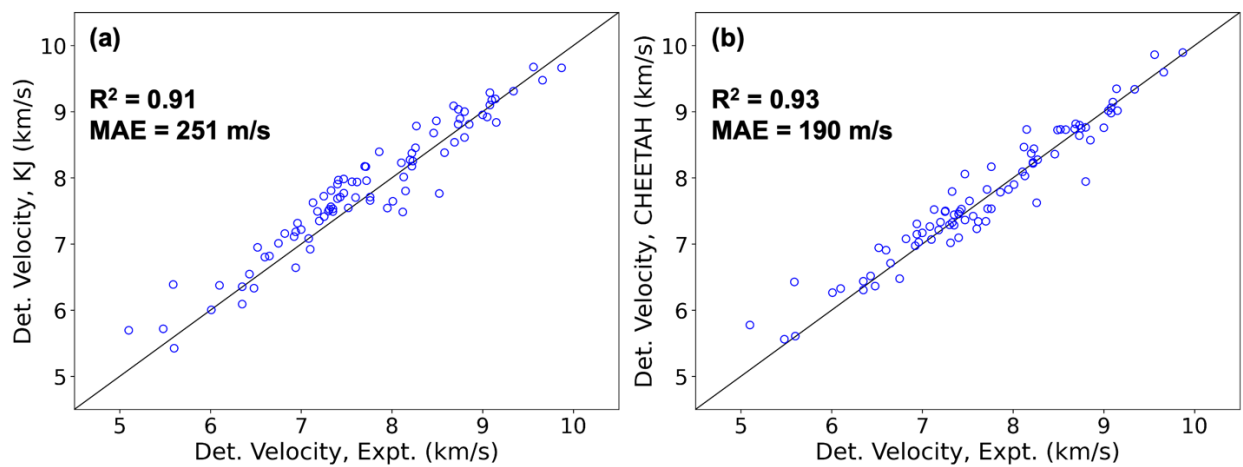

Figure S3 – Comparison of the Kamlet-Jacobs equations (a) and CHEETAH (b) in reproducing experimentally measured detonation velocity using experimental heat of formation values from Muravyey et al.[1]

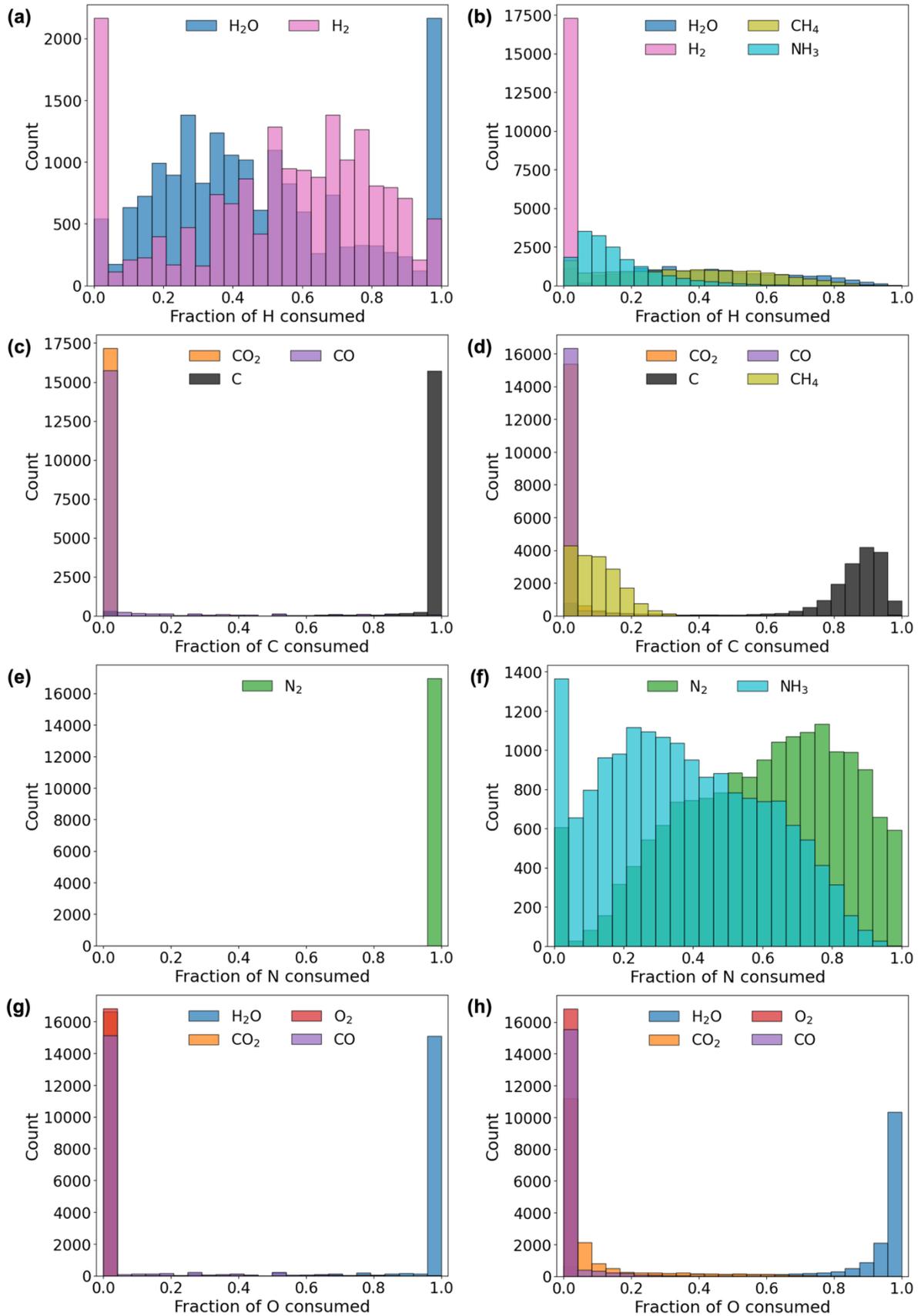

Figure S4 – Amount of each element consumed by various detonation products as predicted by the MKW rules (left column) and CHEETAH (right column). Subplots (a) and (b) correspond to the consumption of H, subplots (c) and (d) correspond to the consumption of C, subplots (e) and (f) correspond to the consumption of N, and subplots (g) and (h) correspond to the consumption of O.

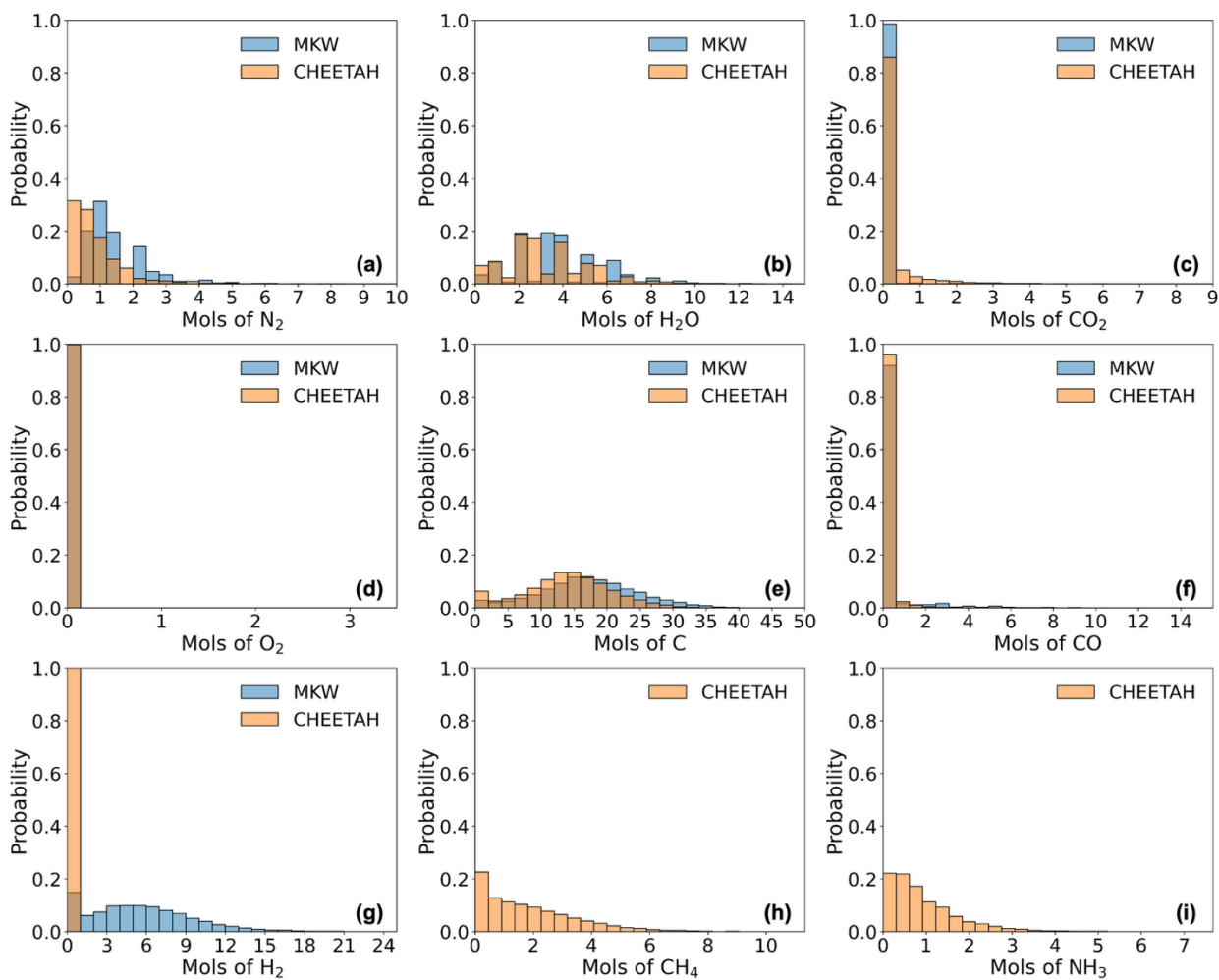

Figure S5 - Distribution of detonation products within the CSD-17k dataset as predicted by the MKW rules and CHEETAH. Each subplot depicts a unique product: (a) $N_2$, (b) $H_2O$, (c) $CO_2$, (d) $O_2$, (e) C (graphite), (f) CO, (g) $H_2$, (h) $CH_4$, and (i) $NH_3$.

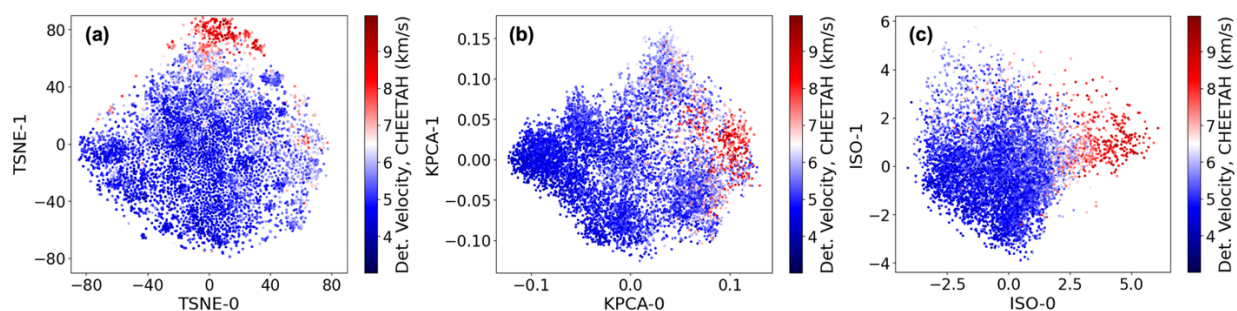

Figure S6 – Morgan fingerprints[3] of the CSD-17k molecules projected into 2D space using t-SNE[4] (a), Kernel PCA[5] (b), and Isomap embedding[6] (c). The color bar depicts the detonation velocity of each molecule as predicted by CHEETAH. Blue points have a low detonation velocity, and red points have a high detonation velocity.

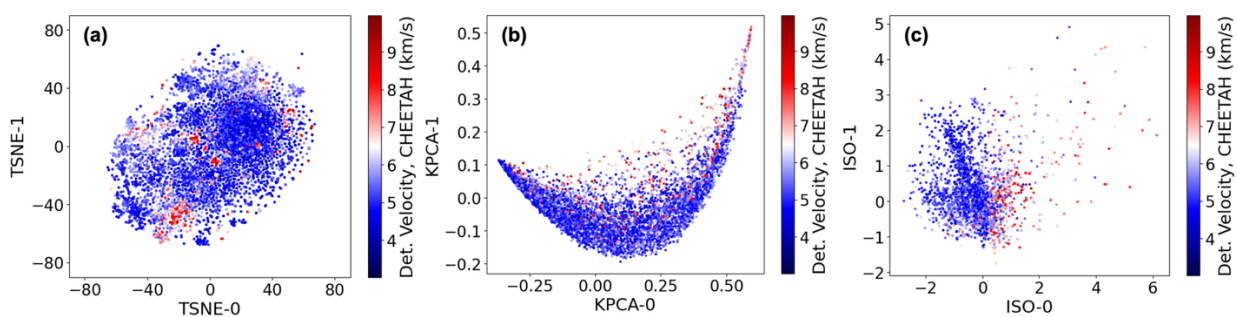

Figure S7 - RDKit fingerprints[7] of the CSD-17k molecules projected into 2D space using t-SNE[4] (a), Kernel PCA[5] (b), and Isomap embedding[6] (c). The color bar depicts the detonation velocity of each molecule as predicted by CHEETAH. Blue points have a low detonation velocity, and red points have a high detonation velocity.

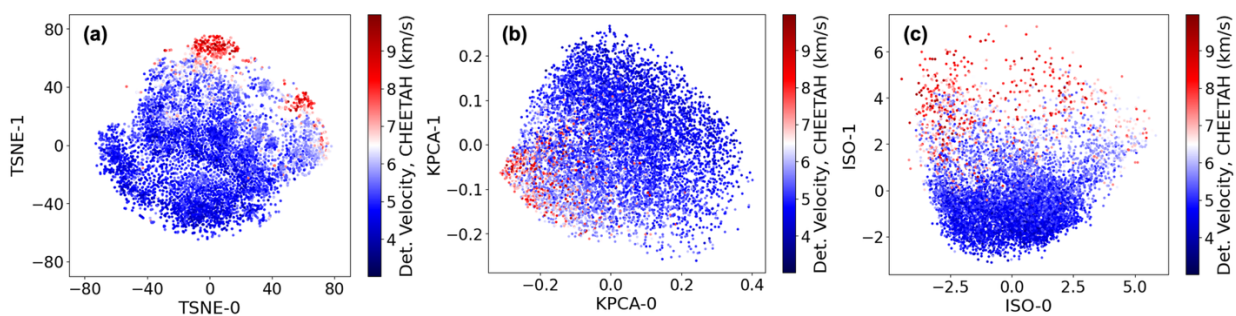

Figure S8 – Atom pair fingerprints[8] of the CSD-17k molecules projected into 2D space using t-SNE[4] (a), Kernel PCA[5] (b), and Isomap embedding[6] (c). The color bar depicts the detonation velocity of each molecule as predicted by CHEETAH. Blue points have a low detonation velocity, and red points have a high detonation velocity.

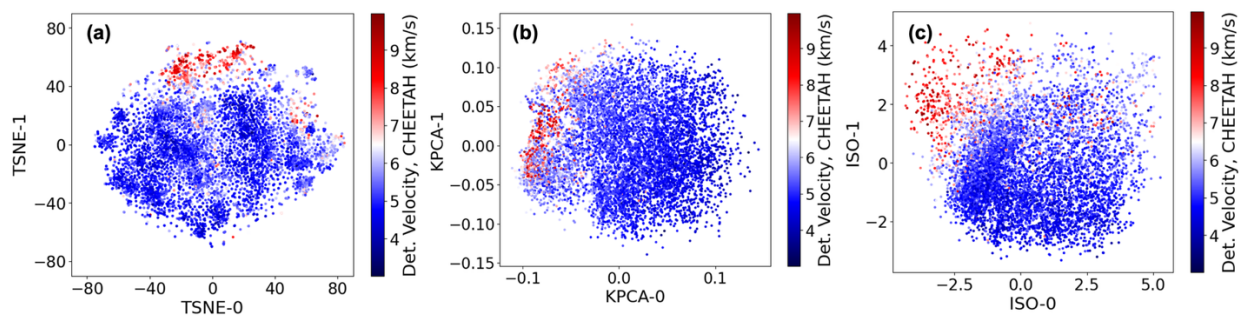

Figure S9 - Topological torsion fingerprints[9] of the CSD-17k molecules projected into 2D space using t-SNE[4] (a), Kernel PCA[5] (b), and Isomap embedding[6] (c). The color bar depicts the detonation velocity of each molecule as predicted by CHEETAH. Blue points have a low detonation velocity, and red points have a high detonation velocity.

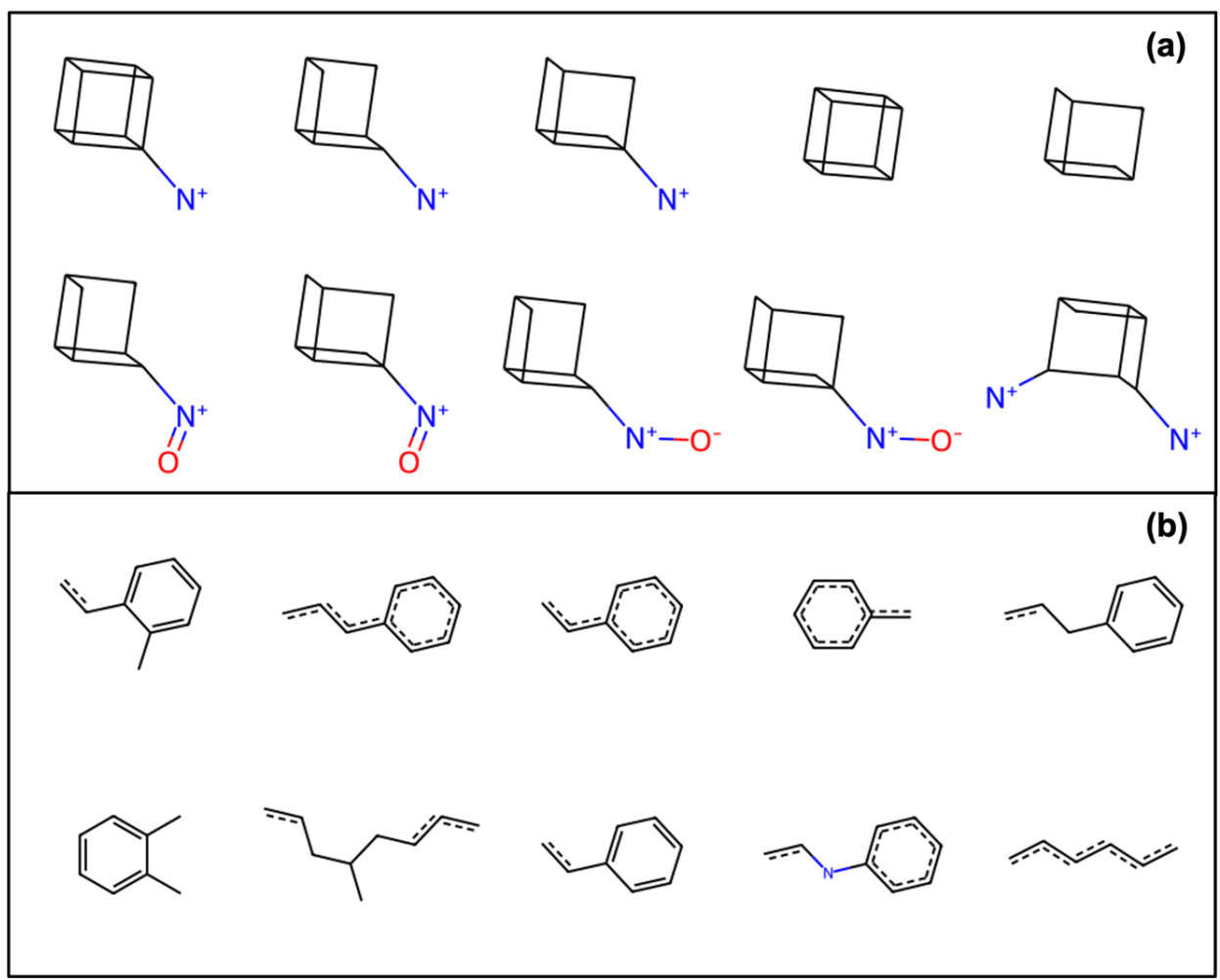

Figure S10 – Principal subgraphs extracted from the (a) top 10% (most performant) and (b) bottom 10% (least performant) of the CSD-17k molecules.

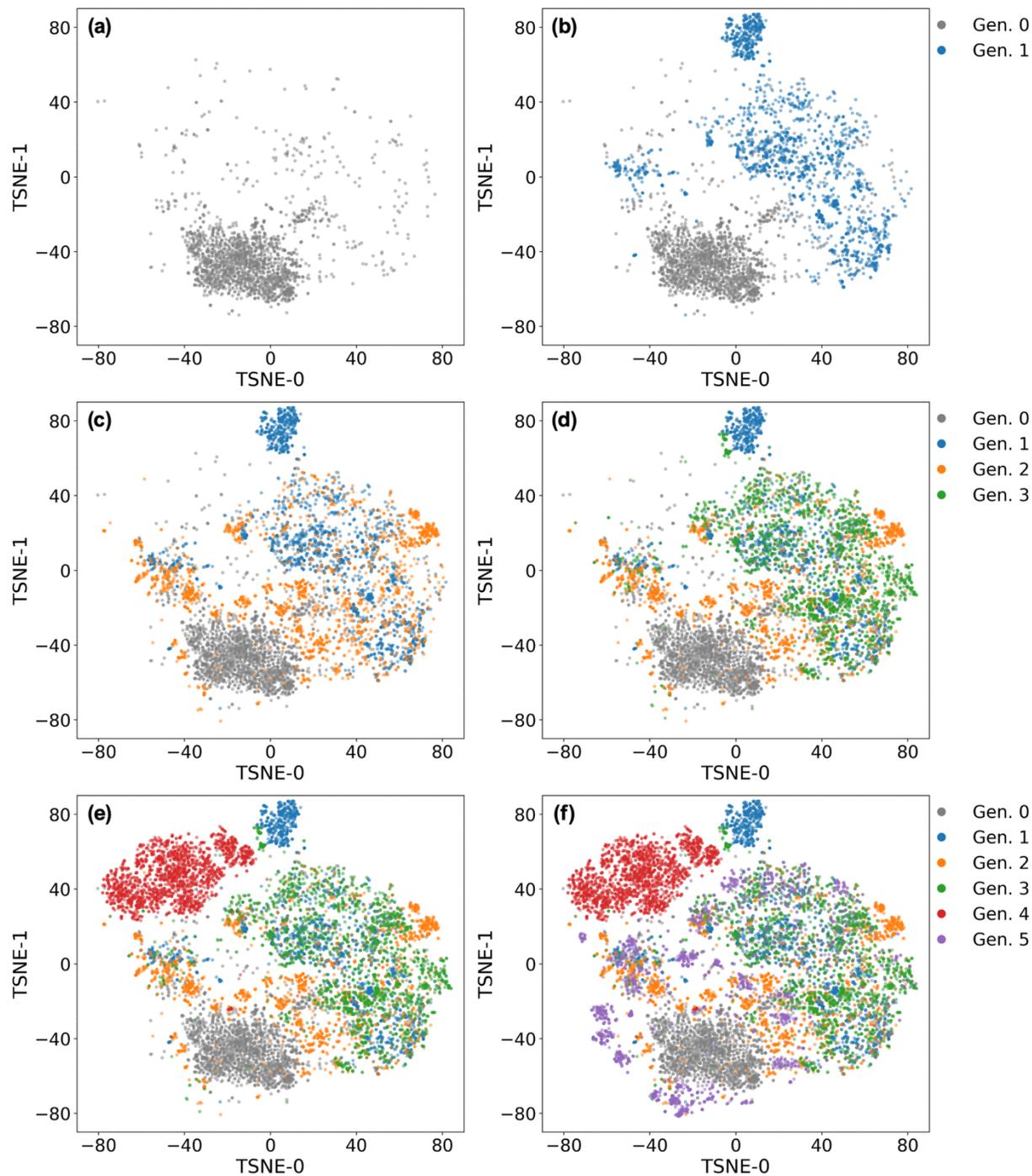

Figure S11 – Chemical space of each active learning generation superimposed in series onto each prior generation. The embedding space is a t-SNE projection of the atom pair fingerprints of each molecule in AL-38k dataset.

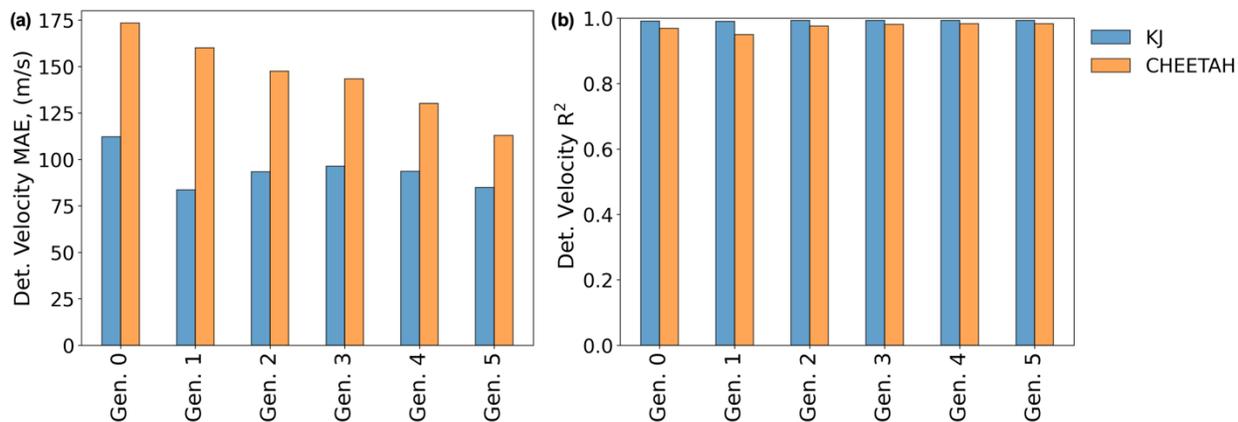

Figure S12 – Predictive performance of the $V_{det}$ surrogate model on the in-fold test set of each active learning generation. Subplots (a) and (b) show the change in MAE and $R^2$ respectively.

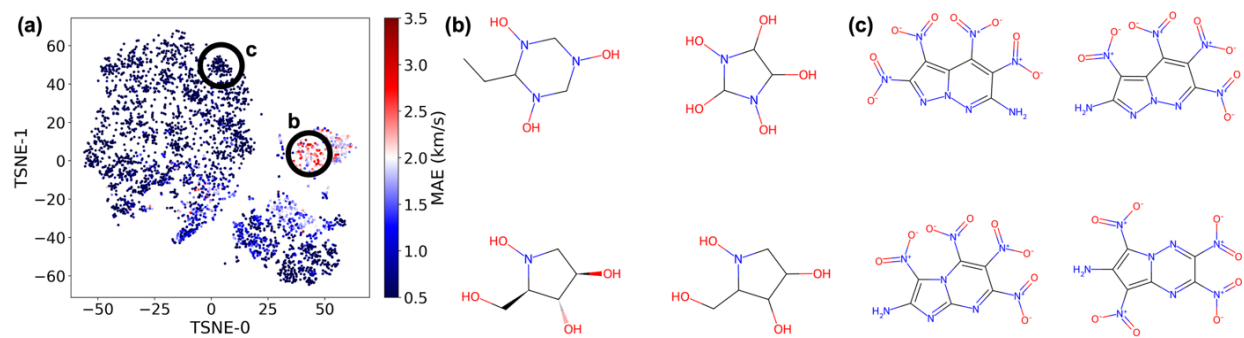

Figure S13 – Representative molecules from active learning Generation 1. Subplot (a) shows a projection of the atom pair fingerprint of each molecule into two-dimensional t-SNE space. Points are color coded by the model's prediction error. Red points indicate molecules with high prediction error and blue points indicate molecules with low prediction error. Subplots (b) and (c) depict atoms taken from the centroid of the corresponding encircled region in subplot (a).

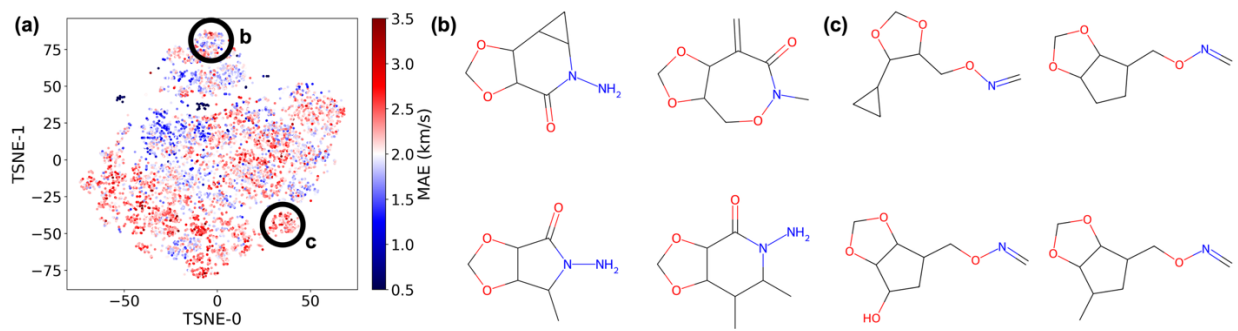

Figure S14 – Representative molecules from active learning Generation 4. Subplot (a) shows a projection of the atom pair fingerprint of each molecule in two-dimensional t-SNE space. Points are color coded by the model's prediction error. Red points indicate molecules with high prediction

error and blue points indicate molecules with low prediction error. Subplots (b) and (c) depict atoms taken from the centroid of the corresponding encircled region in subplot (a).

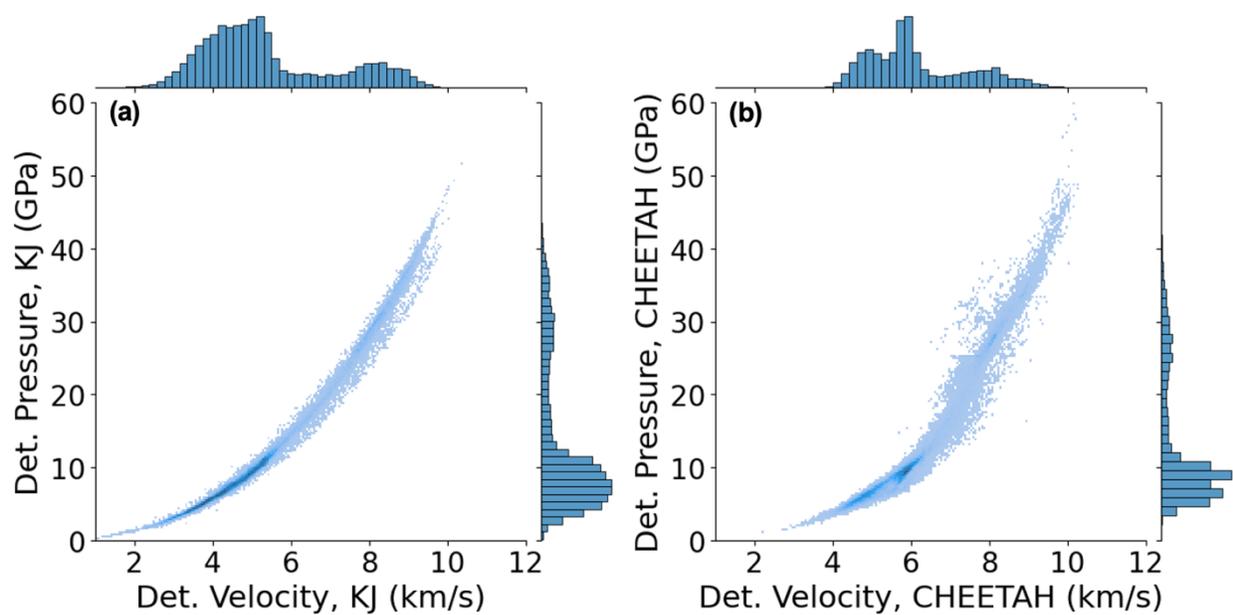

Figure S15 – Detonation performance of all molecules in the AL-38k dataset as predicted by (a) Kamlet-Jacobs and (b) CHEETAH.

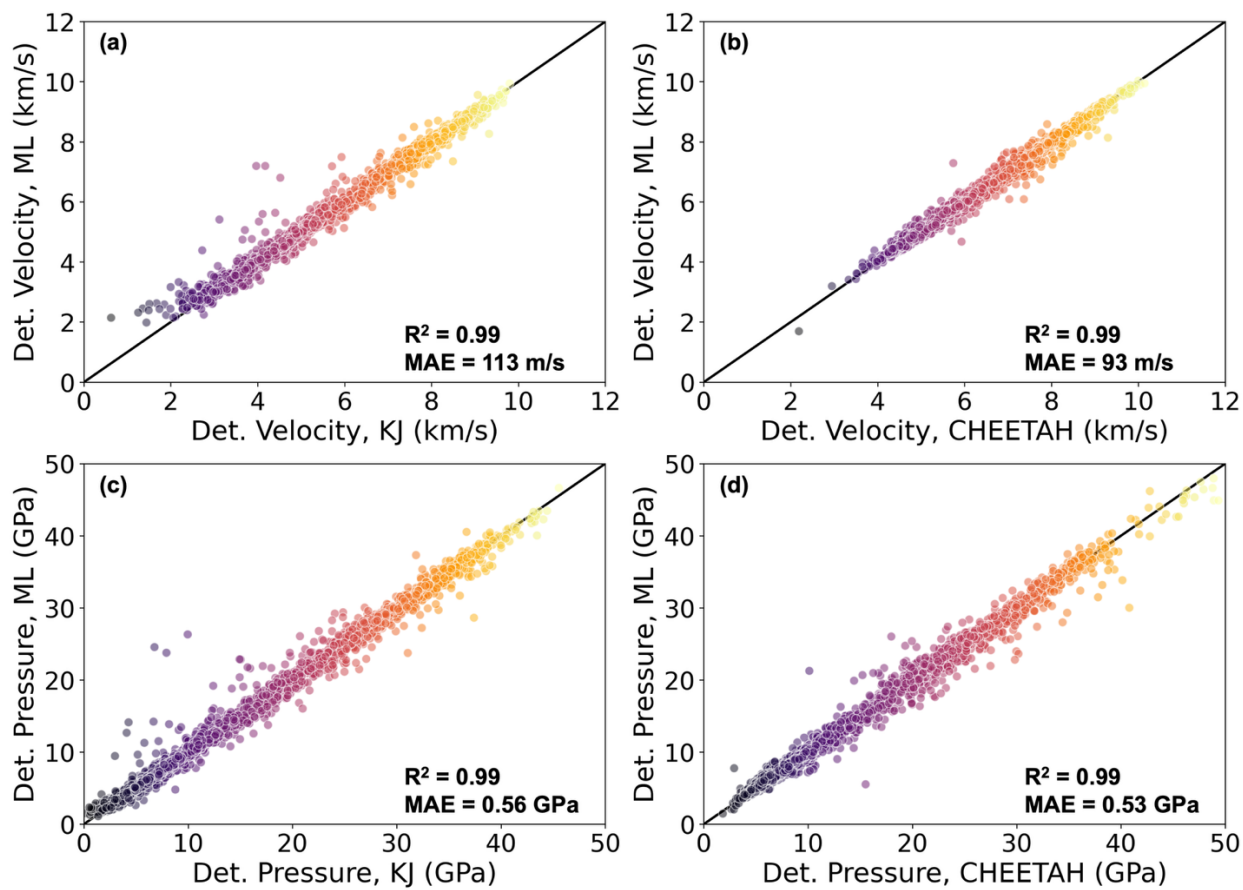

Figure S16 – Parity of the interpretable GBT model in predicting (a) $V_{det}$ as calculated by the Kamlet-Jacobs equations, (b) $V_{det}$ as calculated by CHEETAH, (c) $P_{det}$ as calculated by the Kamlet-Jacobs equations, and (d) $P_{det}$ as calculated by CHEETAH.

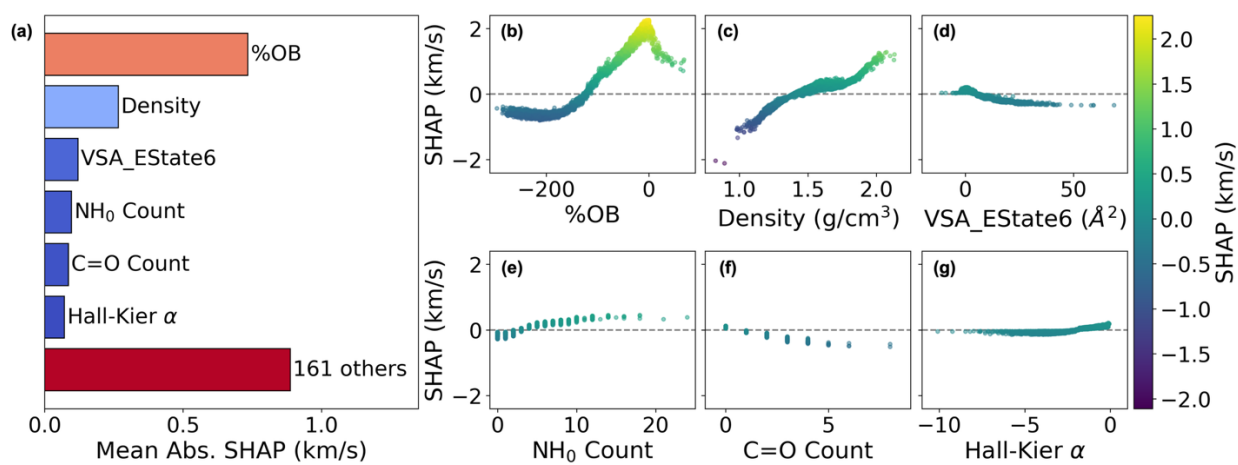

Figure S17 – SHAP importance scores of the top descriptors in predicting $V_{det}$ relative to CHEETAH ground truth. (a) Mean absolute SHAP score for the top six most important descriptors alongside the sum of all other descriptors. (b-g) Scatter plots of SHAP score vs. descriptor value

for each molecule in the GBT model's test set. Each subplot corresponds to a different descriptor: %OB, Density, VSA_EState6, $NH_0$ Count, C=O Count, and Hall-Kier $\alpha$ respectively. The colormap scales linearly with the Y-axis to emphasize data points with the most extreme SHAP scores.

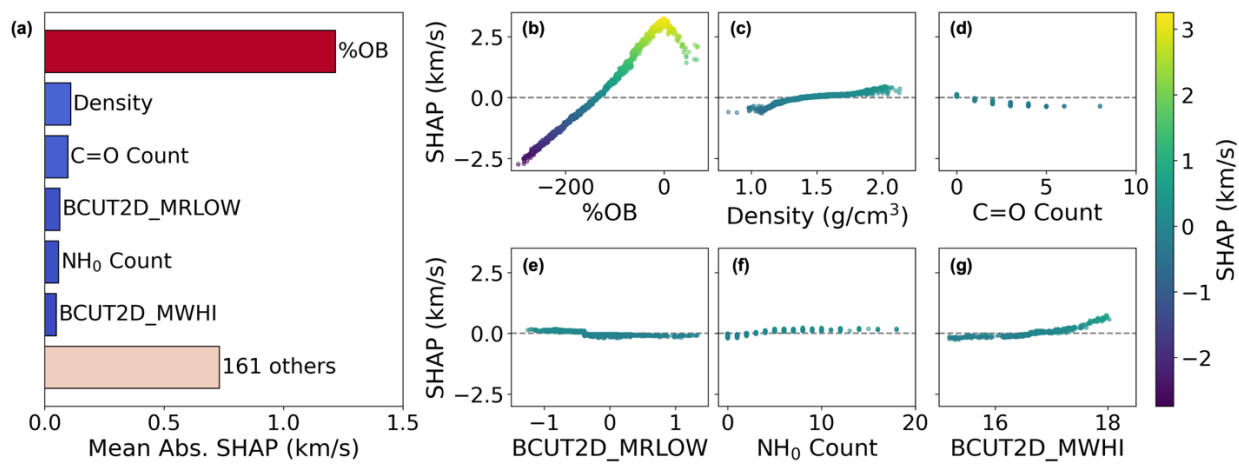

Figure S18 - SHAP importance scores of the top descriptors in predicting $V_{det}$ relative to Kamlet-Jacobs equations ground truth. (a) Mean absolute SHAP score for the top six most important descriptors alongside the sum of all other descriptors. (b-g) Scatter plots of SHAP score vs. descriptor value for each molecule in the GBT model's test set. Each subplot corresponds to a different descriptor: %OB, Density, C=O Count, BCUT2D_MRLOW, $NH_0$ Count, and BCUT2D_MWHI respectively. The colormap scales linearly with the Y-axis to emphasize data points with the most extreme SHAP scores.

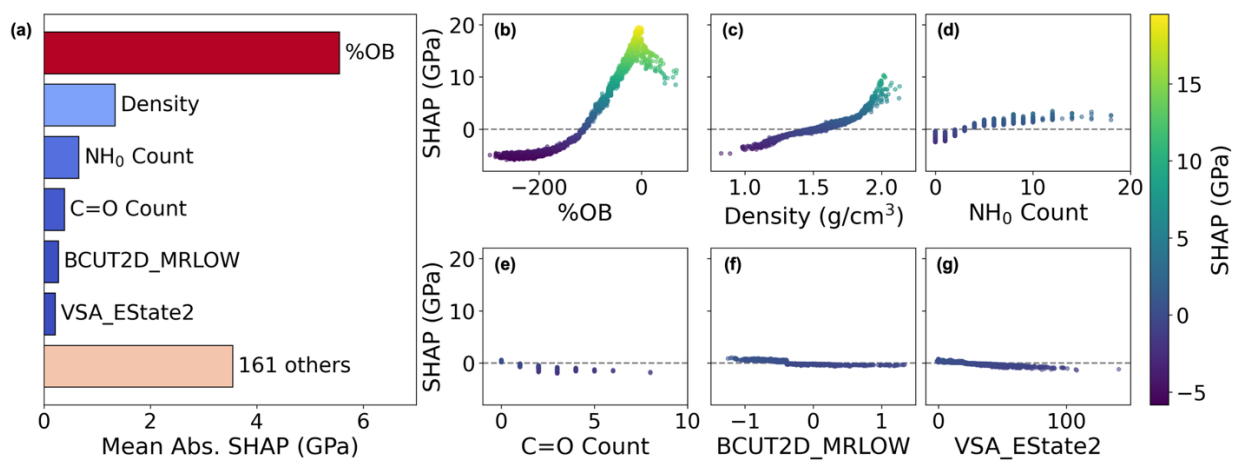

Figure S19 - SHAP importance scores of the top descriptors in predicting $P_{det}$ relative to CHEETAH ground truth. (a) Mean absolute SHAP score for the top six most important descriptors alongside the sum of all other descriptors. (b-g) Scatter plots of SHAP score vs. descriptor value for each molecule in the GBT model's test set. Each subplot corresponds to a different descriptor:

%OB, Density, NH$_0$ Count, C=O Count, BCUT2D_MRLOW, and VSA_EState2 respectively. The colormap scales linearly with the Y-axis to emphasize data points with the most extreme SHAP scores.

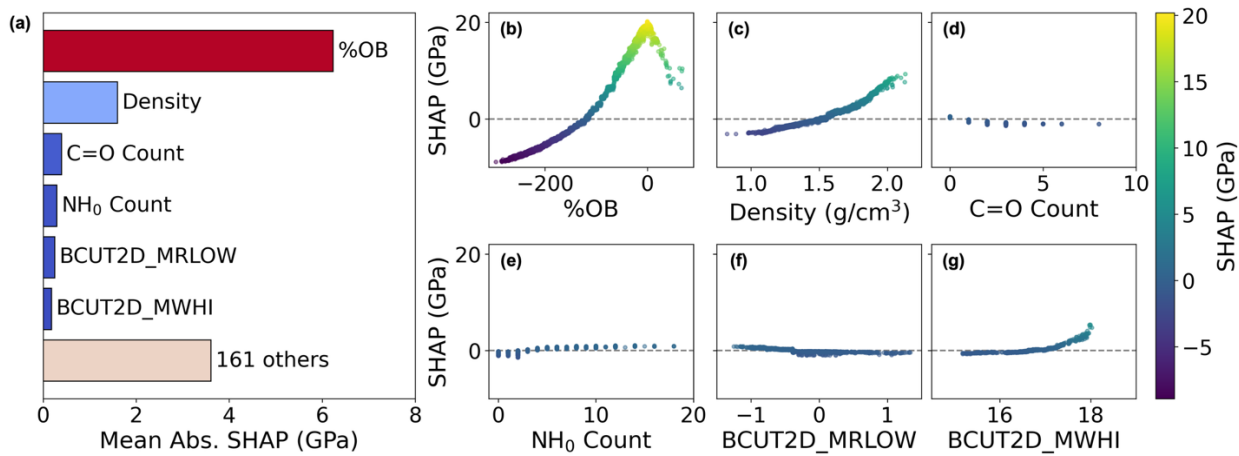

Figure S20 - SHAP importance scores of the top descriptors in predicting $P_{det}$ relative to Kamlet-Jacobs equations ground truth. (a) Mean absolute SHAP score for the top six most important descriptors alongside the sum of all other descriptors. (b-g) Scatter plots of SHAP score vs. descriptor value for each molecule in the GBT model's test set. Each subplot corresponds to a different descriptor: %OB, Density, C=O Count, NH$_0$ Count, BCUT2D_MRLOW, and BCUT2D_MWHI respectively. The colormap scales linearly with the Y-axis to emphasize data points with the most extreme SHAP scores.

Table S1- Grid search parameters used for hyperparameter optimization of the Chemprop surrogate model.

| Hyperparameter | Description | Minimum Value | Maximum Value | Step Size |
|---|---|---|---|---|
| depth | Number of hidden layers. | 2 | 10 | 1 |
| ffn_hidden_dim | Dimensionality of the feed-forward layer. | 300 | 2400 | 50 |
| ffn_num_layers | Number of feed-forward layers. | 1 | 8 | 1 |
| message_hidden_dim | Dimensionality of the message-passing layer. | 300 | 2400 | 50 |
| batch_size | Number of samples to train on per batch. | 16 | 128 | 16 |

Table S2- Grid search parameters for hyperparameter optimization of the GBT feature interpretability model.

| Hyperparameter | Description | Search Space |
|---|---|---|
| learning_rate | Contribution of each tree. | [0.01, 0.1, 1.0] |
| n_estimators | Number of boosting stages to perform. | [300, 600, 1000] |
| max_depth | Maximum depth of the individual regression estimators. | [3, 9, None] |
| min_samples_leaf | Minimum number of samples required to be at a leaf node. | [1, 0.1] |
| min_samples_split | Minimum number of samples required to split an internal node. | [2, 0.1] |
| min_impurity_decrease | Node splitting threshold. | [0.0, 0.1] |
| min_weight_fraction_leaf | Minimum weighted fraction of the sum total of weights required to be at a leaf node. | [0.0, 0.1] |
| max_leaf_nodes | Maximum number of leaves per tree. | [2, 8, None] |

Table S3 – Pairs of highly correlated descriptors. The left column contains the descriptor that was retained for model training, and the right column contains the correlated descriptors that were dropped from consideration.

| Retained Descriptors | Removed Descriptors |
|---|---|
| MaxAbsEStateIndex | MaxEStateIndex |
| MolWt | HeavyAtomMolWt, ExactMolWt, NumValenceElectrons, Chi0, Chi0n, Chi0v, Chi1, Kappa1, LabuteASA, HeavyAtomCount |
| FpDensityMorgan1 | FpDensityMorgan2 |
| Kappa2 | Phi |
| TPSA | NOCount, NumHeteroatoms |
| VSA_EState1 | fr_ether |
| NHOHCount | NumHDonors |
| NumAmideBonds | fr_amide |
| NumAromaticCarbocycles | fr_benzene |
| NumAtomStereoCenters | NumUnspecifiedAtomStereoCenters |
| fr_Al_OH | fr_Al_OH_noTert |
| fr_Ar_NH | Fr_fr_Nhpyrrole |
| fr_Ar_OH | fr_phenol, fr_phenol_noOrthoHbond |
| fr_C=O | fr_C=O_noCOO |

Table S4 - Physical interpretation of each top descriptor as ranked by SHAP values from the interpretable GBT model trained on the AL-38k dataset.

| Descriptor Name | RDKit Name | Meaning |
|---|---|---|
| %OB | - | Oxygen balance. |
| VSA_EState | VSA_EState | Van der Waals surface area contribution from atoms binned by their electrotopological state index. |
| Density | - | Crystalline density. |
| Hall-Kier $\alpha$ | HallKierAlpha | Dimensionless index representing the sum of atom-type correction factors based on covalent radii relative to sp³ carbon. |
| $NH_0$ Count | fr_NH0 | Number of nitrogen atoms with no bonds to a hydrogen. RDKit documentation describes this as tertiary amines, but the pattern also matches $NO_2$ groups. |
| C=O Count | fr_C=O | Number of carbonyl groups. |
| BCUT2D_MRLOW | BCUT2D_MRLOW | Lowest eigenvalue of the Burden matrix weighted by atomic molar refractivity. |
| BCUT2D_MWHI | BCUT2D_MWHI | Highest eigenvalue of the Burden matrix weighted by atomic molecular weight. |